\documentclass[pre,twocolumn,noshowpacs]{revtex4}
\usepackage{epsfig}
\usepackage[]{natbib}
\usepackage{xspace}
\usepackage{color}
\usepackage{bbold}
\usepackage{tabu}
\usepackage{mathtools}

\DeclarePairedDelimiter\bra{\langle}{|}
\DeclarePairedDelimiter\ket{|}{\rangle}
\DeclarePairedDelimiter\bracket{\langle}{\rangle}

\begin{document}

\title{
Quantum walks: the first detected passage time problem}

\author{H. Friedman}
\author{D. A. Kessler}
\author{E. Barkai}
\affiliation{Department of Physics, Institute of Nanotechnology and Advanced Materials, Bar Ilan University, Ramat-Gan
52900, Israel}

\begin{abstract}
Even after decades of research the problem of first passage time statistics
for quantum dynamics remains a challenging 
topic of fundamental and practical importance. 
 Using a projective
measurement approach,  with a sampling time $\tau$, 
we obtain the statistics
of first detection events for quantum dynamics on a lattice, 
with the detector
located at the origin. 
A quantum renewal equation  for 
a first detection wave function, in terms of which the first detection
probability can be calculated,  
 is derived. 
 This formula  
gives the relation between first detection statistics and the solution
of the corresponding Schr\"odinger equation in the absence of measurement.
We demonstrate our results with tight binding quantum walk models. 
We examine a closed system, i.e. a ring, and reveal 
the intricate influence of the sampling time $\tau$
 on the statistics of detection, 
discussing the quantum Zeno effect, half dark states,  revivals
and optimal detection. The initial condition modifies the statistics
of a quantum walk on a finite ring in surprising ways.
In some cases the average detection time is independent of the
sampling time while in others the average exhibits  multiple divergences as
the sampling time is modified. 
For
 an unbounded one dimensional  quantum walk the probability of first detection
decays like $(\mbox{time})^{(-3)}$  
 with  superimposed oscillations, with
exceptional behavior when the sampling period $\tau$ 
times  the tunnelling
rate $\gamma$ is a multiple of  $\pi/2$. The amplitude of the power law decay
is  suppressed
as $\tau\to 0$ due to the Zeno effect.  
Our work presented here, is an extended version of Friedman et al.
arXiv:1603.02046 [cond-mat.stat-mech], and it
  predicts rich physical behaviors
compared 
with classical Brownian motion, 
for which the first passage  probability density decays
monotonically like $(\mbox{time})^{-3/2}$, as elucidated by Schr\"odinger in $1915$.

\end{abstract}
\maketitle

\section{Introduction}

 How long it takes a lion to find its prey,
a particle to reach a domain or
an electric signal to cross a certain threshold? 
These are all examples of the first passage time problem 
\cite{Redner,Hughes,MOR,Benichou}. 
A century ago Schr\"odinger showed  that a Brownian 
particle in one dimension, i.e. the continuous  limit of 
the classical random walk,  starting at $x_0$, will eventually reach 
$x=0$,
with, however, a probability density function (PDF) of the first arrival
time that is fat tailed, in such a way that the mean first passage time diverges
\cite{Schro}.
Ever since, the
 classical first passage time has been a well studied field of research.
 More recently,
much work has been devoted to the analysis of  quantum walks  \cite{Aharonov1,Dorit,Ambainis,Childs,Konno}
(see  \cite{Blumen} for a review). 
These exhibit interference patterns and ballistic scaling and in that
sense exhibit behaviors drastically different from the classical random 
walk. While several variants of quantum walks exist \cite{Blumen},
 for example discrete time
walks, coin tossing walks, and tight-binding models, 
one line of inquiry addresses a question generally
applicable to all these cases, namely
the statistics of first passage or detection times of a quantum particle
(to be defined precisely below).
Quantum walk search algorithms which are
supposed to perform better than classical walk search methods vitalized
this line of research in recent years. 
A physical  example might be the statistics
of the time it takes a single electron, ion or atom 
to reach a detection 
device.
This question, which at first sight appears well-defined and physically
meaningful, has nevertheless been the subject of much controversy.
The Schr\"odinger Eqn. and the standard 
postulates of quantum mechanics \cite{CT} do not give a ready-made 
recipe for calculating these statistics. 
There is no textbook  quantum
operators or wave function associated with the first passage time
measurements
(see \cite{Hauge,Muga,PHD} 
for related historical accounts). Actually, time is a  non-quantum ingredient 
of quantum mechanics and is treated as a object detached from 
the probabilistic interpretation inherent to non-classical reality.
 From the non-deterministic nature of quantum mechanics,
we may expect that the time it takes a single particle
to reach a detection point or domain for a given
Hamiltonian and initial condition should be random 
even in the absence of external noise,
 but how to precisely obtain 
 the distribution of first detection times has remained in our opinion
 a controversial
matter. 

The key to the solution  is that
 we must take into consideration the measurement process
\cite{Bach,Brun06,Brun07,Brun08,Dhar,Dhar1}.
For example consider a zebra sitting at the origin waiting for a lion
to arrive for the first and, unfortunately for her, the last time.
At some rate,  the zebra
records: did the lion arrive or did it not?
 The outcome is a string of answers:
e.g. no, no, no, .... and finally yes.
 If the lion is a quantum particle, then continuous attempts to detect it
by the zebra will maintain
the zebra's life, since the wave function of the lion is collapsed
in the vicinity of the zebra;
this is the famous  quantum Zeno effect \cite{Misra}
 (see more details below).
 On the other hand, if the 
zebra samples the arrival of the lion at a finite constant rate,  
its likelihood of death is much higher.
 In this sense the measurement of the time of  first detection, 
which implies a 
set of null
measurements for times prior to the final positive recording
is very different than  the familiar measurements of canonical variables
like position and momentum. There the system is prepared at time $t=0$ in 
some initial state, it evolves free of measurement until time
$t$, at which point an instantaneous recording of some observable is made.
Furthermore, we must distinguish between first arrival or first
passage  problems \cite{Muga}
and first detection at a site.
Note that even classically the first detection does not imply
that the particle arrived at the site for the first time at the moment of 
detection if the sampling is not continuous in time. 
More importantly,   arrival
times are ill-defined in quantum problems \cite{Aharonov},  because we
cannot have a complete record of the trajectory of a quantum particle,
 whereas  the first detection problem under repeated
stroboscopic measurements is a well-defined problem,  and that is  what we treat in this manuscript.  
 
 Here we investigate the first detection problem of quantum walks
following  Dhar et al. \cite{Dhar,Dhar1} who formulated the problem as a tight-binding
quantum walk, with projective local measurements every $\tau$
units of time (see also \cite{Brun06}). 
 Specifically, we consider 
a particle on a discrete graph, the  quantum evolution determined by the time
independent Hamiltonian $H$. Initially the particle is localized
so the state function is $|\psi \rangle=|x \rangle$ (some of our general results
are not limited to this initial condition, see below).
 Detection
attempts are performed locally at a site we call the origin which is 
denoted with $|0\rangle$. Measurements on the origin are stroboscopic with the sampling time 
$\tau$, and as mentioned the measurement stops once the particle
is detected after $n$ attempts, so $n \tau$ is the  random
first detection time. We investigate 
the statistics of the random observable  $n$.
The questions are: is  the particle eventually detected?
what is the probability of detection after $n$ attempts?
what is the  average of number of attempts of detection before a successful measurement?
This we investigate both for closed systems and open ones. 
 Below we present a physical derivation of the quantum  renewal equation describing
the probability amplitude of first detection for 
the transition $|x \rangle\to |0 \rangle$. In classical stochastic theories
this corresponds to Schr\"odinger's  renewal equation \cite{Schro}
for the first  time  a particle 
starting on $x$ reaches  $0$ (see details below). 
We show
how the solution of Schr\"odinger's wave equation free of measurement
can be used to predict the quantum  statistics of the first detection time.
 Previously Gr\"unbaum et al. \cite{Grunbaum} considered the case where the starting
point is also the detection site $|x\rangle = |0\rangle$.
A topological interpretation of  the detection
process was provided for that initial condition and among other things they showed that
the expected time of first detection is an integer times $\tau$ or infinity.
This integer is the winding number of the so called Schur function of 
the underlying scalar measure, the latter is determined by the initial state and
the unitary dynamics. Hence the expectation of the first detection time is 
quantized \cite{Sinkovicz}. 
 A vastly different behavior is found when we analyse the transition $|x \rangle \to |0 \rangle$ for
$x\ne 0$ \cite{Dhar1,Brun06}. The average of $n$  is not an integer, neither is detection finally
guaranteed. 
 As demonstrated below for a ring geometry half dark states are observed in some cases
while in others the average of $n$  exhibits divergences and non-analytical behaviours,
for certain critical sampling times. Finally, we show that critical sampling,
including slowing down is found even for an infinite system.
Namely, for the quantum walk on the line, 
the first detection probability decays like a power law, with additional
oscillations, where the amplitude of decay is not a continuous function of the sampling rate.
 Thus rich physical behaviors are found for the quantum
first detection problem, if compared with the known results of the classical
random walker.

The spatial  quantum first detection problem is a timely 
subject.  Current
day experiments on quantum walks can be used to study these problems 
in the laboratory \cite{Yaron,QWe2,QWe1,QWe,Xue}. First passage time
statistics in the classical domain 
 are usually recorded  based on single particle analysis.
 Namely, one takes, say, a Brownian
particle, releases it from a certain position and then detects its time of arrival at
some other location.
This single particle experiment is repeated many times and then a 
histogram of the first passage time is reported. While in principle one
could release simultaneously many particles from the same position,
 their mutual interaction will influence the statistics of first arrival
and similarly statistics of identical particles, either bosons or fermions,
 alter
the many particle statistics compared to the single particle case.
Hence, measurement
should be made on single particles, or in other words  at least classically
the first detection 
time is a property of the single particle path and hence its history.
 The recent  advance
of single particle quantum tracking and measurement,
for systems where coherence is maintained for relatively long times, 
is clearly a reason
to be optimistic with respect to possible first detection  measurements.
 Such measurements could test our predictions as well as those
of
a variety of 
other theoretical approaches 
\cite{Muga,Aharonov,Lumpkin,Grot,Stefanak,Shikano,Krap,Ranjith,Das,Miquel},
 some of which are compared with our results 
towards the end of this manuscript.

The navigation map of this  manuscript is as follows.
We  start with the presentation of the  quantum walk model and 
the measurement process in  Sec. \ref{secModel}.
In Sec. \ref{secFDA} the first detection wave function formalism
 is developed.
The main tool for actual solution of the problem is
based on the  generating function formalism
given in  Sec. \ref{secGF}
and in subsection 
\ref{secRENEWAL}
the quantum renewal equation is discussed.  
Sec. \ref{secRings} presents the example of first detection on rings,
with special emphasis on the peculiar statistics 
found on a benzene-like ring.
In Sec. \ref{secFD} we obtain statistics of first detection times,
for a one-dimensional  quantum walk on an infinite lattice.
We end with further discussion of previous results (Sec.
\ref{secOTHER})
 and a summary. 
A short account of part of our main results was recently published
\cite{FKB}.

\section{Model and Basic Formalism}
\label{secModel}

 We consider a particle whose evolution is described by a time independent 
Hermitian Hamiltonian $H$ according to the Schr\"odinger Eqn.
$ i \hbar | \dot{\psi}\rangle = H \left| \psi\right\rangle$.
The initial condition is denoted  $|\psi(0)\rangle$. 
 For simplicity, we consider a discrete $x$-space.
 As an  example we shall later consider the tight binding model
\begin{equation}
H = - \gamma \sum_{x=-\infty}^{\infty} \left(|x\rangle \langle x+1 | + 
|x+1  \rangle \langle x| \right)
\label{eq01aa}
\end{equation}
on a lattice, though our general results are not limited to
a specific Hamiltonian.
 We denote a subset of lattice points $X$, 
and loosely speaking we are interested in
 the statistics of first passage times from the
initial state to any site $x\in X$ in the subset. More generally, $X$ could be any subset of orthogonal states.
An example  is when $X$ 
consists of  a single lattice point, say $x=0$ and initially the
particle is localized at some other lattice 
point $|\psi(0)\rangle = |x'\rangle$. We then investigate the
distribution of the first detection times. For that we must define the 
measurement process following \cite{Dhar,Bach,Ambainis}. 

Measurements on the subset $X$ are made at discrete times
$\tau,2 \tau, \cdots, n \tau \cdots$ and hence
clearly the first  recorded detection
time is either $t_f =\tau$ or $2 \tau$ etc.
The measurement provides two possible outcomes:
either the particle is in $x \in X$ or it is not. 
Consider the first
measurement at time $\tau$. At time $\tau^{-}= \tau - \epsilon$ with 
$\epsilon \to 0$ being positive, the wave function is
\begin{equation}
|\psi(\tau^{-}) \rangle = U(\tau) |\psi(0)\rangle 
\label{eq01abc} 
\end{equation}
and $U(\tau)= \exp( - i H \tau /\hbar)$ as usual. In what follows,
we set $\hbar=1$.
 The probability of finding the
particle in $x\in X$ is, according to the standard interpretation,
\begin{equation}
P_1 = \sum_{x \in X} |\langle x | \psi(\tau^{-})\rangle|^2 .
\label{eq02}
\end{equation}
If the outcome of the measurement is positive, namely the particle is found in $x \in X$,
the first detection time is $t_f=\tau$. On the other hand, if the  particle
is not detected, which occurs with probability $1-P_1$,
 the evolution of the quantum state will
resume. According to collapse theory, following the measurement
 the particle's wave function in $x\in X$ is zero. Namely, a null
measurement alters the wave function in such a way
that  the probability of detecting the
particle in $x \in X$ at time $\tau + \epsilon$ vanishes.
In this sense we are considering projective measurements whose  duration
is very short, while between the measurements the evolution is according to
the  Schr\"odinger Eqn. Mathematically the measurement is a projection \cite{CT},
so that at time $\tau^{+} = \tau + \epsilon$ we have
\begin{equation}
| \psi(\tau^{+}) \rangle = N \left( \mathbb{1} - \sum_{x \in X} | x \rangle \langle x |\right)| \psi(\tau^{-}) \rangle,
\label{eq03}
\end{equation}
where $\mathbb{1}$ is the identity operator, 
and the constant $N$ is determined from the normalization condition.
Here we have used the assumption of a perfect projective measurement
that does not alter  either the relative phases 
or  magnitudes of the wave 
function not interacting with the measurement device, i.e.,
outside the observation domain the wave function is left unchanged beyond a global renormalization. 
This is the fifth postulate of quantum mechanics \cite{CT}, 
though clearly it should be the subject to
continuing  experimental tests.
Since just prior to 
measurement the probability 
of finding the particle in $x$ not belonging to $X$ is $1-P_1$ 
we get
\begin{equation}
\begin{array}{c}
| \psi(\tau^{+}) \rangle = 
{ \mathbb{1} - \sum_{x \in X} |x \rangle \langle x| \over \sqrt{ 1 - P_1}} |\psi(\tau^{-} \rangle = \\
\\
{ \mathbb{ 1} - \sum_{x \in X} |x\rangle \langle x | \over \sqrt{1 - P_1}} U(\tau) |\psi(0)\rangle.
\end{array}
\label{eq04}
\end{equation}
In sum, the measurement nullifies the wave functions on $x\in X$ 
but maintains the relative amplitudes of finding the particles outside
the spatial domain of measurement device,  modifying only the normalization.

 We now proceed in the same way to the second measurement. Between the
first and second detection attempts we have
$ | \psi(2 \tau^{-}) \rangle = U(\tau) |\psi(\tau^{+})\rangle$.
The probability of finding the particle 
 in $x \in X$ at the second measurement,
conditioned  on the quantum walker not having been
 found in the first attempt is
\begin{equation}
P_2= \sum_{x \in X} | \underbrace{\langle x}_{\text{Projection}} | \underbrace{U(\tau)}_{\text{Evolution \ \ }} \underbrace{| \psi(\tau^{+})\rangle}_{\text{  Null\ X \  state} } |^2.
\label{eq05}
\end{equation}
We define the projection operator
\begin{equation}
\hat{D} = \sum_{x \in X} | x \rangle \langle x |
\label{eq06}
\end{equation}
and using Eqs. (\ref{eq04},\ref{eq05}) 
\begin{equation}
P_2 = { \sum_{x \in X} | \langle x | U(\tau)(1-\hat{D}) U(\tau)|\psi(0)\rangle|^2 \over 1- P_1}.
\label{eq07}
\end{equation}
\begin{widetext}
This iteration procedure  is continued to find the
probability of first  detection in the $n$-th measurement, conditioned on
prior measurements  not having detected the particle
\begin{equation}
P_n = \sum_{x \in X} { | \langle x | \left[U(\tau) (1 - \hat{D} )\right]^{n-1} U(\tau) | \psi(0)\rangle|^2 \over (1 - P_1 ) .... (1- P_{n-1}) }. 
\label{eq08}
\end{equation}
\end{widetext} 
In the numerator the operator  $1 - \hat{D}$ appears  $n-1$ times 
corresponding to the $n-1$ prior measurements. Similarly, in the 
denominator we find
$n-1$ probabilities of null measurements $1-P_j$.  
Following \cite{Dhar,Dhar1} we define the first detection wave function
\begin{equation}
|\theta_n \rangle = U(\tau) \left[ \left( 1 - \hat{D} \right) U\left( \tau \right) \right]^{n-1} |\psi(0)\rangle
\label{eq09}
\end{equation}
or equivalently 
$|\theta_n \rangle=[U(\tau)(1-\hat{D})]^{n-1}|\theta_1\rangle$ 
with the initial condition
$|\theta_1\rangle = U(\tau)| \psi(0)\rangle$. 
The bra $|\theta_n\rangle$ is defined only for the moments of detection
$n=1,2,\cdots$,
unlike $|\psi(t)\rangle$ which is a function of continuous time.  
With this definition 
\begin{equation}
P_n = { \langle \theta_n| \hat{D} | \theta_n \rangle \over \Pi_{j=1} ^{n-1} (1 - P_i) } .
\label{eq10}
\end{equation}
%


The main focus of this work is on the probability of first detection
in the  $n$-th measurement, denoted $F_n$. This is of course not the same as
$P_n$ which as mentioned is a conditional probability, namely the 
 probability of detection on
 the $n$-th attempt given no previous detection.
The conceptual measurement  process  for the calculation
of $F_n$ is as follows. We start with an initial spatial wave function 
$|\psi(0)\rangle$ and evolve it until time $\tau$ when the
detection of the particle in $x\in X$ is attempted, and  with probability 
$P_1$ the first measurement is also the first detection. Hence to simulate this process  on a
computer we toss a coin using an uniform random number generator and if 
the particle is detected the measurement time is $\tau$. 
 If the particle
is not detected we compute $P_2$. 
Then at time $2 \tau$  either the particle is 
detected with probability $P_2$
 or not. 
This process is repeated  until a measurement is recorded (see remark below)
 and that 
measurement constitutes the random  first detection event. 
 In order to gain statistics of the first detection time we
return to the initial step and restart the process with the 
same initial condition. 
In this way, repeating this many times,
 we construct the 
first detection probability 
\begin{equation}
F_n = (1-P_1)(1-P_2)... (1 - P_{n-1}) P_n .
\label{eq11}
\end{equation}
Using Eq. (\ref{eq10}) we obtain
\begin{equation}
F_n = \langle \theta_n | \hat{D} | \theta_n \rangle .
\label{eq12}
\end{equation}
We see that the first detection probability $F_n$ is the expectation value
of the projection  operator $\hat{D}$ 
with respect to 
$|\theta_n\rangle$, which we term 
 the first detection wave function.

{\bf Remark.} We shall see that not all sequences of
measurements, generated on a computer or in the lab,
 yield a detection  in the long-time limit. 
This is not problematic since also
classical random walks in say three dimensions are not recurrent and hence
the total 
probability of detection is not necessarily  unity. 
In many works one defines the survival probability, i.e.,
 the probability that
the particle is not detected
in the first $n$ attempts,
\begin{equation}
S_n = 1- \sum_{n=1}^n F_n.
\label{eqSur}
\end{equation}
 The eventual  survival probability 
$S_\infty$ 
can be equal zero or not. If the initial condition and the detection location
are identical and $S_\infty=0$ the quantum walk is called recurrent.  
We will later investigate whether or not the 
quantum walk is recurrent, both  for the cases of an infinite lattice and 
a finite ring.

\section{First detection amplitude}
\label{secFDA}

In this section, we solve the first detection time problem
 for quantum dynamics
with a single detection site, which we label $x=0$, so
$\hat{D} = | 0 \rangle \langle 0|$.
 We define the amplitude of the first detection
as
\begin{equation}
\phi_n= \langle 0| \theta_n \rangle 
\label{eq13}
\end{equation}
so that $F_n = |\phi_n|^2$.
Using Eq. (\ref{eq09})  $\phi_1= \langle 0| U(\tau) | \psi(0)\rangle$, 
$\phi_2= \langle 0 | U( 2 \tau)|\psi(0)\rangle - \phi_1 \langle 0 | U(\tau) | 0\rangle$ and a short calculation yields
\begin{equation}
\phi_3 = \langle 0| U(3 \tau) | \psi(0)\rangle -\phi_1 \langle 0| U(2 \tau) |0\rangle  - \phi_2 \langle 0| U(\tau)|0\rangle.
\label{eq14}
\end{equation}
In Appendix \ref{AppITGM}, using induction we find our first main equation
\begin{equation}
\phi_n = \langle 0 | U(n \tau) | \psi(0)\rangle -
\sum_{j=1} ^{n-1} \phi_j \langle 0 | U\left[\left(n-j\right) \tau\right] | 0 \rangle.
\label{eq15}
\end{equation}
We call this iteration rule
 the quantum renewal equation. 
It yields the amplitude $\phi_n$ in terms 
of a propagation free of measurement; i.e., 
$ \langle 0 | U(n \tau) | \psi(0)\rangle $ is the amplitude for being at
 the origin at time $n \tau$ in the absence of measurements, from which we
subtract $n-1$ terms related to the previous history of the system.
The physical interpretation of Eq. (\ref{eq15})
 is
that the condition of non-detection in previous measurements translates into
subtracting  wave sources (hence the minus sign)
 at the detection site $|0\rangle$ following the $j$th detection attempt.
This is due to the nullification of the wave function at the
 detection site
in the $j$th  measurement. 
 The evolution of that wave source from the $j$th 
measurement onward  is described by the free Hamiltonian, hence
$\langle 0| U[(n-j)\tau]|0\rangle$ which gives the amplitude of return back
to the origin, in the time interval $(j \tau,n \tau)$.

We now consider the formal solution to the first detection problem for 
an initial condition on the origin hence $|\psi(0)\rangle=|0\rangle$
and 
as mentioned the origin is also the point at which we perform the detection trials.  
Clearly in this case $\phi_1=\langle 0|U(\tau)|0\rangle$ and since
$U(0)=\mathbb{1}$ we get $\phi_1=1$ when $\tau \rightarrow 0$ which is
expected. For $\phi_2= \langle 0|U(2)|0\rangle - \langle 0|U(1)0\rangle^2$
where we use the short-hand notation
$U(n)\equiv U(n \tau)$.
Similarly
$\phi_3=\langle 0| U(3)   |0\rangle
-2\langle 0| U(2) |0\rangle \langle 0| U(1) |0\rangle +
\langle 0| U(1)   |0\rangle^3$. 
 The general solution is obtained by iteration
using Eq. 
(\ref{eq15}),
\begin{equation}
\phi_n = \sum_{i=1} ^n \sum_{\{m_1,\cdots,m_i\}} (-1)^{i+1} \langle 0| U(m_1)|0\rangle \cdots \langle 0| U(m_i) |0 \rangle.
\label{eq16}
\end{equation}
The double  sum is over all partitions of $n$, i.e. all
 $i$-tuples of positive integers 
$\{m_1, ....m_i\}$
satisfying $m_1+\cdots+m_i=n$. For example for $n=5$ we have five 
partitions 
corresponding to $i=1, \cdots, 5$, for $i=1$ the set in the second sum is
 $\{ 5 \}$, for $i=2$ we sum over $\{2,3\},\{3,2\},\{1,4\}$ and $\{4,1\}$,
for $i=3$ we use  $\{1,1,3\},\{1,3,1\},\{3,1,1\},\{2,2,1\},\{2,1,2\},\{1,2,2\}$,
for $i=4$,  $\{1,1,1,2\},\{1,1,2,1\},\{1,2,1,1\},\{2,1,1,1\}$
and for $i=5$ we have one term $\{1,1,1,1,1\}$.
Hence 
\begin{equation}
\begin{array}{l}
\phi_5 = \langle 0|U(5)|0\rangle- 2 \langle 0 | U(4) | 0\rangle\langle 0|U(1) |0\rangle+ \\
3 \langle 0|U(1)|0\rangle^2 \langle 0 |U(3) |0\rangle-
4 \langle 0|U(1) |0\rangle^3 \langle 0|U(2)|0\rangle +\\
3 \langle 0 | U(2) | 0\rangle^2 \langle 0| U(1) | 0 \rangle-
2 \langle 0 | U(3)| 0  \rangle \langle 0 | U(2) | 0 \rangle +\\
\langle 0 | U(1) | 0\rangle^5.
\end{array}
\label{eq17}
\end{equation}
With a 
symbolic  program
like Mathematica one can obtain similar  exact expressions for intermediate
values of $n$. However, to gain some insight
 we
turn now to the generating function approach \cite{Brown}.

\section{Generating function Approach} 
\label{secGF}

The $Z$ transform,  or discrete Laplace transform, 
of $\phi_n$ is by definition \cite{Brown,remark1}
\begin{equation}
\hat{\phi}(z) = \sum_{n=1}^\infty z^n  \phi_n.
\label{eq18}
\end{equation}
$\hat{\phi}(z)$ is also called the generating function. 
Multiplying  Eq. (\ref{eq15}) by $z^n$ and summing over $n$
\begin{multline}
\hat{\phi}(z) = \sum_{n=1}^\infty \langle 0 | z^n U(n) |\psi(0)\rangle - \\
\sum_{n=1} ^\infty \sum_{j=1}^{n-1} \phi_j z^j \langle 0| z^{n-j} U(n-j)|0\rangle.
\label{eq19}
\end{multline}
Evaluating  the first term on the right hand side we get
\begin{equation}
\hat{U}(z) = \sum_{n=1} ^\infty z^n U(n) = \sum_{n=1} ^\infty \exp( - i H \tau n) z^n={ z e^{ - i H \tau } \over 1 - z e^{ - i H \tau}}.
\label{eq20}
\end{equation} 
The second term in Eq. (\ref{eq19}) is a convolution term and after
rearrangement we find one of our main results \cite{remark1}
\begin{equation}
\hat{\phi}(z) = { \langle 0 | \hat{U}(z) | \psi(0)\rangle \over 1 + \langle 0 | \hat{U}(z) |0 \rangle}
\label{eq21}
\end{equation}
or more explicitly
\begin{equation}
\hat{\phi}(z) =  {\langle 0| { 1  \over z^{-1}e^{i H \tau} - 1}  | \psi(0)\rangle \over 1  + \langle 0 | {1  \over z^{-1} e^{i H \tau}  - 1} | 0 \rangle}.
\label{eq22}
\end{equation}
This equation, relates  the generating function $\hat{\phi}(z)$ to
the Hamiltonian
evolution between the initial condition and the detection attempt. 

This approach is also valid for
other types of  measurements, repeatedly performed at times
 $\tau,2 \tau, \cdots$.
For example the case where we  measure a set of points $x \in X$ is
given in Eq. 
(\ref{eq:general_z_transform})
in  Appendix 
\ref{AppITGM}.
 First detection  measurements
of general observables is also treated there.

\subsubsection{Relations between $\hat{\phi}(z)$ and
$\phi_n$, $S_\infty$ and  $\langle n \rangle$.}

As usual the amplitudes $\phi_n$ are given in terms of their
$Z$-transforms by the inversion formula
\begin{equation}
\phi_n = { 1 \over  n !} { d^n \over d z^n} \hat{\phi}(z)|_{z=0}
\label{eq22pn}
\end{equation}
or
\begin{equation}
\phi_n = { 1 \over 2 \pi i} \oint_C \hat{\phi}(z) z^{ - n -1} {\rm d} z
\label{eq23}
\end{equation}
where $C$ is a counter clockwise path that contains the origin and is
entirely within the radius of convergence of $\hat{\phi}(z)$. 

The probability of being measured 
is also related to the generating function $\hat{\phi}(z)$ by
$$ 1-S_\infty=  \sum_{n=1} ^\infty F_n = \sum_{n=1} ^\infty |\phi_n|^2 = $$
\begin{equation}
{1 \over 2 \pi} \int_0 ^{2 \pi} \sum_{k=1} ^\infty \phi_k  e^{ i \theta k} 
\sum_{l=1} ^\infty  \phi_l ^{*}  e^{ - i \theta l} {\rm d} \theta=
{1\over 2 \pi} \int_0 ^{2 \pi} |\hat{\phi}(e^{i \theta})|^2 {\rm d} \theta.
\label{eqTot}
\end{equation}
Similarly 
\begin{equation}
\langle n \rangle = \sum_{n=1} ^\infty n F_n = 
{1 \over 2 \pi} \int_0 ^{2 \pi} \left[\hat{\phi}\left(e^{i \theta}\right)\right]^{*}
\left(- i {\partial \over \partial \theta} \right)\hat{\phi}(e^{i \theta}) {\rm d} \theta.
\label{eqAve}
\end{equation}
The latter is the average of $n$ only when the particle
is detected with probability one, namely when  $S_\infty=0$. 
A shorthand notation of Eq. (\ref{eqAve})
is 
$\langle n \rangle = \langle \hat{\phi}| -i \partial_{\theta} | \hat{\phi}\rangle$.

\subsection{Connection between first detection and spatial wave function}
\label{secRENEWAL}

 In classical random walk theory the key approach to the first passage time
problem is to relate it to occupation probabilities \cite{Redner}.
Let us unravel a similar relation in the quantum domain, connecting
between first detection statistics and the corresponding  
wave packet, namely the time dependent
solution of the Schr\"odinger equation in the absence
of measurement  (see also
\cite{Grunbaum} for the $|0\rangle \rightarrow |0\rangle$ transition). 
To that end,  we first  briefly review the classical random walk. 

Consider a 
classical random walk in discrete time
 $t=0,1,..$, 
for example 
a random walk on a cubic lattice in dimension $d$ 
with jumps to nearest neighbours. The main assumption is that
the random walk is Markovian.
Denote $P_{\rm cl}({\bf r},t)$ as the probability
that the walker
 is at ${\bf r}$ at time $t$ when initially the particle is at the origin
${\bf r} = 0$
and in the absence of any absorption. 
Let $F_{\rm cl}({\bf r}, t)$ be the first passage probability:
 the probability 
that the random walk visits site ${\bf r}$ for the first time 
at time $t$ with the same initial condition.
Following  the first equation in
the  first chapter in \cite{Redner}, $P_{{\rm cl}}({\bf r}, t)$
and $F_{\rm cl}({\bf r},t)$ are related  
by
\begin{equation}
P_{\rm cl}({\bf r}, t) = \delta_{{\bf r} 0} \delta_{t 0} +
\sum_{t'\le t} F_{\rm cl}( {\bf r}, t') P_{{\rm cl}}(0,t-t').
\label{eqRed01}
\end{equation}
This equation \cite{Schro,Montroll,MontrollW},
 sometimes called the renewal equation, is generally valid
for Markov processes in the sense
that it is not limited to discrete time and space models; in the  continuum
one needs only to replace summation with integration, 
and probabilities by probability densities. 
The idea 
behind Eq. (\ref{eqRed01}) 
is that a particle on position ${\bf r}$ at time $t$ must have either arrived
there previously at time $t'$ {\em for the first time}
and then returned back or it arrived
at ${\bf r}$ exactly at time $t$ for the first time (the $t'=t$ term)
\cite{Schro,Redner}.  
Using the $Z$ transform the following equations are derived \cite{Redner}
\begin{equation}
F_{\rm cl}( {\bf r} , z) = \left\{
\begin{array}{c l}
{P_{\rm cl}( {\bf r}, z) \over P_{\rm cl}(0, z) } & {\bf r} \neq 0 \\
1- {1  \over P_{\rm cl}(0, z) } \quad & {\bf r} = 0. 
\end{array}
\right.
\label{eqRed02}
\end{equation}
From this formula, various basic properties of random walks
 can be derived. One example is
 the P\'olya theorem which
answers the question: does a particle eventually return to its origin;
i.e., whether the random walk is recurrent.
 A second
is that in  one dimension, for an open system
without bias, the famous law  $F_{\rm cl} (0,t) \sim t^{-3/2}$ is found
for large first passage time $t$  and hence the first
passage time has an infinite mean, as mentioned in the introduction.
We will later find the quantum analogue to this well known $t^{-3/2}$ behaviour.

 At first glance this classical picture might not seem related to ours.
However  consider the case where we detect the particle at the origin,
so $\hat{D}=|0\rangle \langle 0|$ and  initially 
 $|\psi(0)\rangle=|0\rangle$.
Then Eq. (\ref{eq22}) reads
\begin{equation}
\hat{\phi}(z) = { \langle 0 | { 1 \over z^{-1} e^{i H \tau} -1} |0 \rangle
\over 1 + 
 \langle 0 | { 1 \over z^{-1} e^{i H \tau} -1} |0 \rangle}.
\label{eqQuRed02}
\end{equation}
We add and subtract one in the numerator and use
$\langle 0 | 0 \rangle=1$
and 
\begin{equation}
1 + {1 \over z^{-1} e^{i H \tau} - 1} = { 1 \over 1 - z e^{- i H \tau} } 
\label{eqIdd}
\end{equation}
 to rewrite 
Eq. (\ref{eqQuRed02})
 as
\begin{equation}
\hat{\phi}(z) = 1 - {1 \over \langle 0  | {1 \over 1 - z e^{-i H \tau}} | 0 \rangle},
\label{eqRed03}
\end{equation}
Expanding in $z$ we get  a geometric series 
\begin{equation}
\hat{\phi}(z) = 1 - { 1 \over \langle 0 | \sum_{n=0} ^\infty z^n \exp( - i H \tau n) | 0 \rangle}.
\label{eqRed04}
\end{equation}
By definition the sum 
$\langle 0|\sum_{n=0} ^\infty z^n \exp( - i H \tau n) | 0 \rangle$
 is the generating 
function
of the amplitude of being at the origin
retrieved from the solution of the Schr\"odinger equation without detection. 
Namely, let $|\psi_f(t)\rangle$ be the solution of the Schr\"odinger
equation for the same initial condition $|\psi_f(0)\rangle =|0\rangle$
(the subscript $f$ denotes a wave function free of measurement). 
The amplitude of being at the
origin at time $t$  is
 $\langle 0|\psi_f(t)\rangle$ and 
$|\psi_f(t)\rangle = \exp( - i H t) |0 \rangle$ as usual.
 We define the generating function 
 of this amplitude, for the sequence of measurements under consideration
\begin{equation}
\langle 0 | \hat{\psi}_f(z)\rangle_0  \equiv  \sum_{n=0} ^\infty z^n \langle 0 | \psi_f( n \tau)\rangle 
\label{eqRed05}
\end{equation}
and clearly $\langle 0 | \hat{\psi}_f(z)\rangle_0 =\sum_{n=0} ^\infty \langle 0 | z^n \exp( - i H \tau n) | 0 \rangle$, the subscript zero 
denotes the initial condition. 
Hence we get the appealing result
reported already in \cite{Grunbaum}
\begin{equation}
\hat{\phi}(z) = 1 - {1 \over \langle 0 | \hat{\psi}_f(z)\rangle_0} .
\label{eqRed06}
\end{equation}
Thus the generating function of the first detection time
is determined from the $Z$ transform
of the  spatial wave function at the point of detection $x=0$. 
 This connection is  the quantum  analogue of the second
line in the classical expression Eq. (\ref{eqRed02}) 
since in both cases  we start and detect at the origin. 

 Similarly for an initial condition initially localized at some site $x\neq 0$,
so $| \psi(0)\rangle=|x\rangle$ with detection at site $0$ we find
\begin{equation}
\hat{\phi}(z) = {\langle 0 |\hat{\psi}_f(z)\rangle_x \over \langle 0 | \hat{\psi}_f(z)\rangle_0}
\label{eqRed06aa}
\end{equation}
where $|\hat{\psi}_f(z)\rangle_x$ is the $Z$ transform of the wave function
free of measurements initially localized on site $x$,
$|\hat{\psi}_f(z)\rangle_x=\sum_{n=0} ^\infty z^n |\psi_f(n\tau)\rangle_x$ 
with
$|\psi_f(n \tau) \rangle_x = \exp(- i H n \tau)|x \rangle$. 
We see that the ratio of the generating functions of the amplitudes
of finding the particle on $|0\rangle$ for initial condition on $x$ 
and the location of measurement site $0$, 
obtained from the measurement-free evolution, yields the generating 
function of the measurement process. This is the sought after quantum renewal 
equation, namely the amplitude analogue of the upper line of the classical
Eq.  
(\ref{eqRed02}).

{\bf Remark} 
In Eq. 
(\ref{eqRed05}) 
the lower limit of the sum is $n=0$, while
in Eq. (\ref{eq18}) the sum starts at $n=1$ as noted already \cite{remark1}.
Since $\phi_0=0$ one may of course use a summation from $0$ also in Eq.
(\ref{eq18}). 

{\bf Remark} Our formalism is not limited to spatially
 homogeneous Hamiltonians. Note that in our classical discussion,
following the textbook treatment~\cite{Redner} and
for the sake of simplicity, we have assumed translation
invariant random walks. In non-translation invariant systems,
one should replace $P_{{\rm cl}}(0,t-t')$ in the left hand side
of  Eq.
(\ref{eqRed01})
with 
$P_{{\rm cl}}({\bf r} , t- t'| {\bf r} ,0)$. 
Since the convolution structure of the equation remains,
related to the Markovian
hypothesis, Eq. 
(\ref{eqRed02})
can be easily modified to include non-homogeneous effects. 

{\bf Remark} Sinkovicz et al. \cite{Sinkovicz1}
 found a quantum Kac-Lemma for recurrence
time, thus analogies between quantum and classical walks are not limited
to the renewal equation under investigation.

\subsection{Zeno Effect}

 As pointed out in \cite{Brun08,Dhar}
 when $\tau \rightarrow 0$ we find the Zeno effect
\cite{Misra,Patil}.
Since in that limit $\exp(-i H \tau) =1$ and $\hat{U}(z)=z/(1-z)$, 
using Eq. 
(\ref{eq21}) we get
\begin{equation}
\lim_{\tau\rightarrow 0} \hat{\phi}(z) = z \langle 0 |\psi(0)\rangle.
\label{eq27}
\end{equation}
The amplitude of finding the particle
at the origin in the first attempt, is given by the  initial wave function
projected on the origin, i.e. the probability amplitude of finding
the particle at the origin at $t=0$.
 Hence the above expression gives an obvious answer
for the first measurement;  the repeated measurements being very frequent
do not allow the wave function to be built up at the origin, and hence 
$\phi_n=0$ for all $n>1$.
 This means that we may investigate the problem
for  $\tau$ small relative to the time scales of the Hamiltonian, but we cannot
take the limit $\tau \rightarrow 0$ if we wish to retain  information 
on the measurement process beyond
the initial state. 

\subsection{Energy representation}

Eq.  
(\ref{eq20})
for a time-independent Hamiltonian
yields
\begin{equation}
\langle E_m | \hat{U}(z) | E_i \rangle
 = \left[ z^{-1} \exp\left(  i E_m \tau\right) -1\right]^{-1} \delta_{mi}
\label{eq25}
\end{equation}
 so that the operator $\hat{U}(z)$ is diagonal in the energy representation.  
Here $|E_i \rangle$ is a stationary state of the Hamiltonian $H$,
namely $H|E_i\rangle = E_i |E_i \rangle$. 
Clearly it is worthwhile presenting the solution in that basis.
Consider the example of the  measurement at the spatial origin corresponding
to state $|0\rangle$. This state can be expanded in the energy representation
$|0\rangle = \sum_{k} C_k | E_k \rangle$ with $C_k=\langle E_k | 0 \rangle$. 
Here as usual $\langle E_m | E_k \rangle = \delta_{mk}$. 
Similarly the initial condition is expanded
as $|\psi(0)\rangle = \sum_{k} A_k| E_k\rangle$. The matrix
element 
\begin{equation}
\langle 0| \hat{U}(z) | \psi(0)\rangle =  
\sum_k C_k ^{*} A_k\left[z^{-1}  \exp(i E_k \tau)-1\right]^{-1} 
\label{eq28}
\end{equation}
together with
\begin{equation}
\langle 0 | \hat{U}(z) | 0 \rangle =
\sum_k  |C_k|^2 \left[z^{-1}  \exp(i E_k \tau)-1\right]^{-1} 
\label{eq29}
\end{equation}
yields $\hat{\phi}(z)$ using Eq. (\ref{eq21}). 
For the special case where $|\psi(0)\rangle = |0\rangle$ we get $A_k=C_k$ and 
\begin{equation}
\hat{\phi}(z) = {\sum_k |C_k|^2 \left[ z^{-1} \exp(i E_k \tau) -1\right]^{-1} 
\over
1+ \sum_k |C_k|^2 \left[ z^{-1} \exp(i E_k \tau) -1\right]^{-1} }
\label{eq30}
\end{equation}
Here as usual $\sum_k |C_k|^2 = 1$.
It is easy to check that when $\tau\rightarrow 0$ we get $F_1=|\phi_1|^2 =1$
since a particle starting at the origin is with probability one detected when
$\tau \rightarrow 0$. 

\begin{figure}
\vspace*{-0.3in}
\begin{center}
\includegraphics[width=0.4\textwidth]{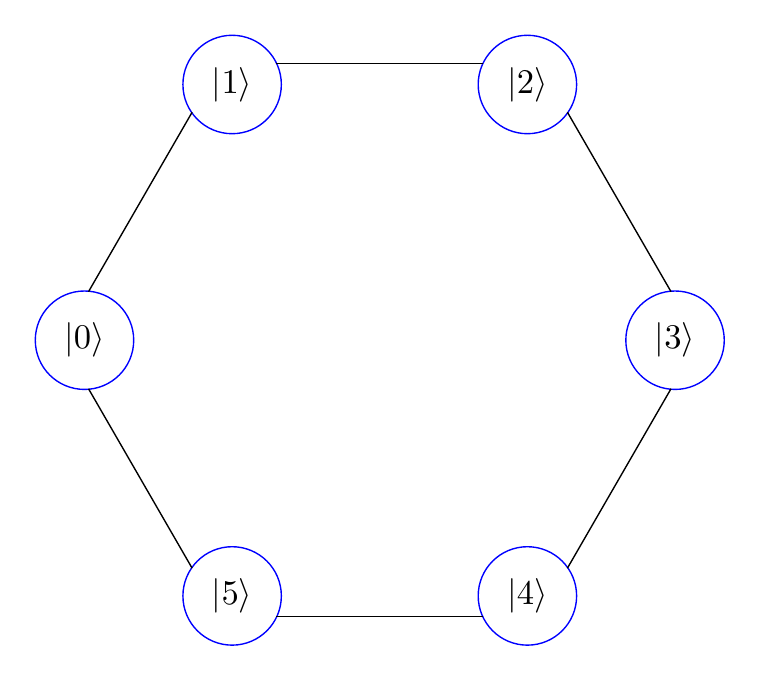}
\end{center}
\vspace*{-0.2in}
\caption{
Schematic model of a benzene ring. 
In the text measurement is performed on site $0$ and we discuss
several initial conditions. 
}
\label{figBenzene}
\end{figure}

\begin{figure}
\centering
\includegraphics[width=0.40\textwidth]{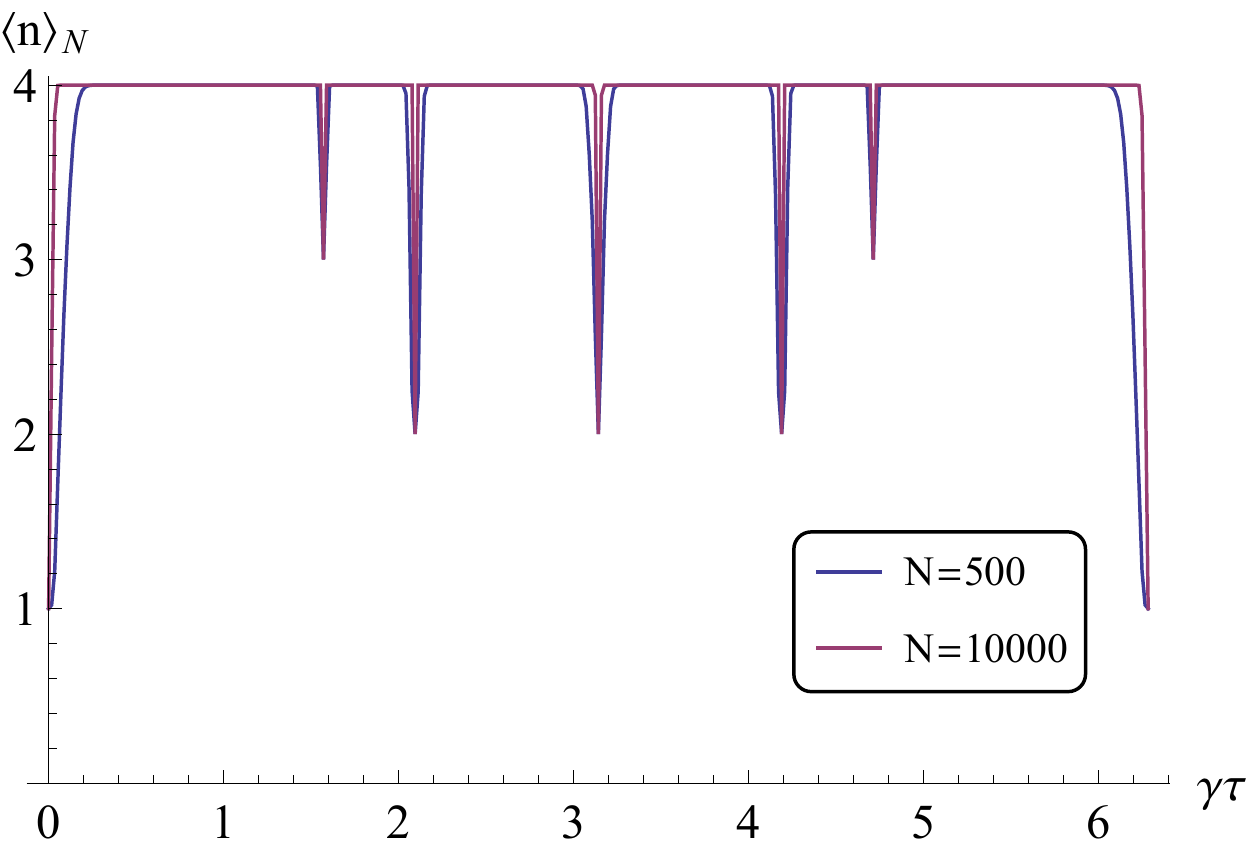}
\caption{
For a  quantum walk on a benzene ring, with initial condition 
$|\psi(0)\rangle= |0\rangle$ and projective measurements on the origin,
the average number of detection attempts is, by Eq. (\ref{eqfour}), $4$ except for 
the cases when $\gamma \tau$ is a multiple of $\pi/2$ or $2 \pi/3$.
 Here we plot $\langle n \rangle_N =\sum_{n=1} ^N n F_n$ for $N = 500$
and $N=10000$,
the results converging (not shown) as we increase $N$ further.   
}
\label{fig2four}
\end{figure}

\section{Rings}
\label{secRings}
For our explicit calculations, we will focus
on tight binding models in one dimension~\cite{Blumen}. The first model
is a quantum walk on a ring of length $L$:
\begin{equation}
H = - \gamma \sum_{x=0}^{L-1} \left(|x\rangle \langle x+1 | + 
|x+1  \rangle \langle x| \right).
\label{eqTBring}
\end{equation}
This  describes a quantum particle jumping between nearest neighbours
on the ring. We use periodic boundary conditions and thus from the site
 labeled $x=L-1$ one may 
jump either to the origin $x=0$ or to the site labeled $x=L-2$. 
In condensed matter physics the parameter $\gamma$ is called the
tunnelling rate. 

\subsection{Benzene-Type Ring}

 As our first example we consider the tight-binding model on a hexagonal ring
presented in Fig. 
\ref{figBenzene},
namely a structure similar to the benzene molecule \cite{CT,Feynman}. 
We consider the influence of initial states $|\psi(0)\rangle=|x\rangle$ with
$x=0,1,..5$ 
on the statistics of first detection times for detection at site $0$
so $\hat{D} = |0\rangle\langle 0 |$. 
According to our theory, to find the generating function we need
the energy levels of $H$ and its eigenstates.    
 The six energy levels of the system
 are $E_k= - 2 \gamma \cos(\theta_k)$
with $k=0,...5$ and the eigenstates are $|E_k\rangle^{T}=(1,e^{i \theta_k},e^{2 i \theta_k},e^{i 3 \theta_k} , e^{ i 4 \theta_k},e^{i 5 \theta_k})/\sqrt{6}$
with $\theta_k=2 \pi k/6$ \cite{CT} ($\ ^T$ is the transpose). 
Hence the coefficients 
 $|C_k|^2=|\langle E_k|0\rangle|^2=1/6$, reflecting the symmetry of the problem.

\subsubsection{Starting at $x=0$}

We use Eq. (\ref{eq30}) and find
\begin{equation}
\hat{\phi}(z) = {{ 1 \over 6} \sum_{k=0} ^5 { 1 \over z^{-1} \exp( i E_k \tau)-1}
\over 
1 + {1 \over 6} \sum_{k=0} ^5 { 1 \over z^{-1} \exp( i E_k \tau) -1}}.
\label{eq31}
\end{equation} 
The nondegenerate energy levels  are $-2 \gamma$ and $2 \gamma$ 
while $-\gamma$ and $\gamma$ are doubly degenerate,  hence
for real $z$
\begin{equation}
\hat{\phi}(z)=
{ 
{1 \over 3} \left\{ \mbox{Re} \left[ { 1 \over z^{-1} e^{2 i \gamma \tau} -1} \right] +
2 \mbox{Re} \left[ {1 \over z^{-1} e^{i \gamma \tau} -1} \right] \right\}
\over 
1+  {1 \over 3} \left\{ \mbox{Re} \left[ { 1 \over z^{-1} e^{2 i \gamma \tau} -1} \right] +
2 \mbox{Re} \left[ {1 \over z^{-1} e^{i \gamma \tau} -1} \right] \right\}
}.
\label{eqDel}
\end{equation}
It is interesting to note that
the generating function  satisfies the identity
\begin{equation}
\hat{\phi}(e^{i\theta})\hat{\phi}(e^{- i \theta}) = 1. 
\label{ident1}
\end{equation}
an  identity we will return to below when discussing $\langle n \rangle$
and $S_\infty$.
Inserting Eq. (\ref{ident1}) in Eq. 
(\ref{eqTot}) and integrating over $\theta$  gives $S_\infty=0$.
Thus the survival probability is zero in the long time
limit.
This  behavior is classical in the sense that
for  finite systems a classical random walker is always detected.  
Note that for a quantum walker this conclusion is not generally valid.
 If we start at $|1\rangle$ for example and measure at $|0\rangle$, 
and perform measurements on full revival periods, 
the particle is never recorded (see further details
and other examples  below). 
Hence, for a quantum particle
 the survival
probability $S_n$ does not generally decay to zero as $n\to\infty$, even for finite systems.  

 For special values of $\gamma \tau$ we get exceptional
behaviors.
When $\gamma \tau$ is $2 \pi$ times  an integer we get
$\hat{\phi}(z)=z$ namely the 
 measurement in the  first attempt is made with probability $1$,
so the first detection time is $\tau$, which is expected
since the wave function is fully revived at these $\tau$'s
 in its initial state at the origin
\cite{Blumen}.
If $\gamma \tau=\pi$ we get $\hat{\phi}(z) = z(3 z-1)/(3-z)$. Inverting
we find $\phi_1=-1/3$ and $\phi_n =8/3^n $ for $n \ge 2$,
thus the amplitude $\phi_n$ decays exponentially.
It follows that the first detection probabilities are
\begin{equation}
F_n = \left\{ 
\begin{array}{c c}
{1 \over 9} \ \ \ \ \ \ n=1 \\
 \ \ \\
{64 \over 9^n} \ \ \ \ \  n \ge 2 .
\end{array}
\right.
 \label{eq31b}
\end{equation}
The average number of detection attempts is $\sum_{n=1} ^\infty n F_n = 2$.
If $\gamma \tau =  \pi/2$ we find $\hat{\phi}(z)= -z(1+2 z  + 3z^2)/ (3 + 2z + z^2)$
 which has
simple poles and hence 
\begin{equation}
F_n = \left\{
\begin{array}{c c}
{ 1/9}
 \ \ \ \ \ \ n=1 \\
{ 16/81}
 \ \ \ \ \ \ n=2 \\
{24\over3^n}\sin^2{\big(
\zeta_1 (n-2)
-
\zeta_2 \big)}
 \ \ \ \ \ \ n\ge3
\end{array}
\right.
 \label{eq31baa}
\end{equation}
where
$\zeta_1=\tan^{-1}(\sqrt{2})$ and
$\zeta_2=\tan^{-1}(\sqrt{2}/5)$. 
For this case $\langle n \rangle = 3$. Similarly for $\gamma \tau=2 \pi/3 + 2 k \pi$ 
and $\gamma \tau = 4 \pi/3+ 2 k \pi$ we
get $\langle n \rangle = 2$. 
The general feature of finite rings is an exponential decay of 
$F_n$  with a superimposed oscillation determined by the poles of 
the generating function.  However the sampling times
$\gamma \tau=0,\pi/2, 2 \pi/3, \pi\cdots$  considered so far
 exhibit behaviors
which are not typical, as we now show.

A surprising behavior is found for the average, with
\begin{equation}
\langle n \rangle = 4 
\label{eqfour}
\end{equation}
for any sampling rate in the interval $(0, 2 \pi)$ besides
what we call the exceptional sampling times
  $\gamma \tau=0,\pi/2,2 \pi/3 ,\pi, \cdots $ where as mentioned
$\langle n \rangle=1,3,2,2, \cdots$ respectively,
 which is continued periodically (see Fig. \ref{fig2four}). This result is derived below.
 As mentioned in the introduction the fact that $\langle n \rangle$ is some integer was
already pointed out rather generally by \cite{Grunbaum} and this is related to topological
effects. 
Except for the exceptional points, the variance of $n$ is
\begin{widetext}
\begin{equation}
\mbox{Var}(n) = -11 + {27\over 4 - 4 \cos \gamma \tau} + {1\over
6 \cos^2 \gamma \tau} + {3\over 4  + 4\cos \gamma \tau} + {3\over
(1 + 2 \cos\gamma \tau)^2}
\end{equation}
\end{widetext}
 so that the first detection time exhibits large fluctuations near these 
 points.
Thus for the $0$ to $0$ transition
it is only the average $\langle n \rangle$ that is nearly always
 not sensitive to the sampling
rate, not the full distribution of first detection times.

There are numerous methods to  find $\langle n \rangle =\sum_{n=1} ^\infty
n F_n$. 
 For the exceptional points we used the exact solution for $F_n$
 (as mentioned).
For other sampling times we use two approaches: the first using
Mathematica and is based on a Taylor expansion
of $\hat{\phi}(z)$  and the second is an analytic calculation.
The former approach is very general in the sense that it can be used
in principle for general initial conditions and other problems beyond the benzene ring.

Specifically, we calculate $F_n$ exactly using the expansion
of $\hat{\phi}(z)$ with  
  symbolic programming on Mathematica. This is performed
up to some large $N$. We then calculate $\langle n \rangle_N = \sum_{n=1} ^N 
n F_n$. Clearly $\langle n \rangle > \langle n \rangle_N$, and  increasing $N$
we see convergence towards $\langle n \rangle= 4$ except for the mentioned
exceptional points. An example is shown in Fig. 
(\ref{fig2four}) for the cases $N= 500$ and $N=10000$. 

Even better is to write $\hat{\phi}(z)= z^4 H(1/z) /H(z)$, which is the extension to general $z$ of the identity
Eq. (\ref{ident1})
that we used to show $S_\infty=0$.
To find $\langle n \rangle$ we use Eq.
(\ref{eq31}) to find
\begin{multline}
H(z) = [2\cos(\gamma\tau)+\cos(2\gamma\tau)]z^3-\\
[3+6\cos(\gamma\tau)\cos(2\gamma\tau)]z^2+ \\
[4\cos(\gamma\tau)+5\cos(2\gamma\tau)]z-3, 
\label{eqnn}
\end{multline}
and with Eq. (\ref{eqAve})
\begin{multline}
\langle n \rangle= \\
\frac{1}{2\pi i}\int_0^{2\pi}\frac{e^{-4i\theta}H(e^{i\theta})}{H(e^{-i\theta})}
\frac{\partial}{\partial \theta}
\left(\frac{e^{4i\theta}H(e^{-i\theta})}{H(e^{i\theta})}\right)d\theta \\
=4-\frac{1}{i\pi}\int_0^{2\pi}\frac{\partial}{\partial \theta}\ln{H(e^{i\theta})}d\theta.
\end{multline}
Rewriting  $H(z)=a(z-z_1)(z-z_2)(z-z_3)$ we can proceed to
show that
\begin{equation}
\langle n \rangle=
4-\frac{1}{i \pi}\sum_{j=1}^{3}\int_{0}^{2\pi}\frac{\partial}{\partial \theta}\ln(e^{i\theta}-z_j)d\theta=4-2\alpha-\beta
\end{equation}
where $\alpha$ (or $\beta$) is the number of zeros of  $H(z)$  for $z$ within
(or on)  the unit  
circle  respectively.  As explained in Appendix B, 
 $\alpha=0$ for otherwise we would find $F_n>1$.
For the exceptional values of $\gamma\tau$ we find $\beta>0$, as follows:
\begin{equation}
\beta=
\begin{cases}
1&\gamma\tau=\frac{1}{2}\pi+\pi k\\
2&\gamma\tau=\frac{2}{3}\pi+2\pi k, \pi+2\pi k,\frac{4}{3}\pi+2\pi k\\
3&\gamma\tau=2\pi k\\
0&\text{otherwise}.
\end{cases}
\end{equation}
This agrees with the values of $\langle n \rangle$ 
we have found at the exceptional 
points. This exercise shows that mathematically, at least for
this example,  the exceptional
points are those specific
values of $\tau$ where some of the zeros of the polynomial $H(z)$ are found  to lie on the unit circle in the complex plane.
 We will soon find a by far more physical 
and explicit formula for these points, Eq. (\ref{eqET}) below.

$$ $$
\begin{table}[h!]
\begin{tabular}{ c | c | c | c | c | c | c | c | c}
\centering
 $x$ & $0<\gamma \tau < 2 \pi ^{*}$  &$ \gamma\tau=  0$ & $\frac{1}{2}\pi$ & $\frac{2}{3}\pi$ & $\pi$ & $\frac{4}{3}\pi$ & $\frac{3}{2}\pi$ & $2\pi$ \\
 \hline
  0 &  1  & 1 &  1  & 1 &  1  & 1 &  1  & 1\\
  1 & 1/2 & 0 & 1/6 & 0 &  0  & 0 & 1/6 & 0\\
  2 & 1/2 & 0 & 1/2 & 0 & 1/2 & 0 & 1/2 & 0\\
  3 &  1  & 0 & 2/3 & 1 &  0  & 1 & 2/3 & 0\\
\end{tabular}
\caption{Total detection probability $1- S_\infty$ for a quantum walker
on a benzene
ring, for different localized
 starting points $|\psi(0)\rangle = | x \rangle$. Measurements are performed at $x=0$ hence
initial conditions on sites $1$ and $2$ are equivalent to initial conditions
on $5$ and $4$ respectively. Values of the parameter
 $\gamma \tau$ are listed in
the first row, and $0<\gamma \tau<2 \pi ^{*}$ implies all values of 
$\gamma \tau$ in the interval, 
 besides the listed special cases, e.g. $\gamma\tau=\pi$.  
}
\label{TableExcpetional}
\end{table}
 
\subsubsection{Half Dark states}

 Another peculiar behavior is found if the detection is at
the origin  $\hat{D} = |0\rangle\langle 0 |$ and the starting point is 
$|i\rangle$ with $i=1,2,4,5$.
The total probability of detection is found to be, by the method explained in Appendix B,
\begin{equation}
1-S_\infty=1/2
\label{eqHalf}
\end{equation}
 for all
values of $0<\gamma \tau < 2 \pi$ besides exceptional 
points which are listed in Table 
\ref{TableExcpetional}. 
The exceptions include the case  when $\tau$ is the
full revival time, for which case the probability of being detected
is of course $0$. 
The behaviour  Eq. (\ref{eqHalf})  was observed in
\cite{Dhar,Dhar1} for even larger systems. 
It is  remarkable that  for certain initial conditions, the detection
of the particle is not guaranteed, and only in half of the measurement
processes we detect the particle, hence we call these initial
conditions
half  dark 
states. 

\subsubsection{Starting on site $3$ measuring on $0$}

\begin{figure}
\centering
\includegraphics[width=0.49\textwidth]{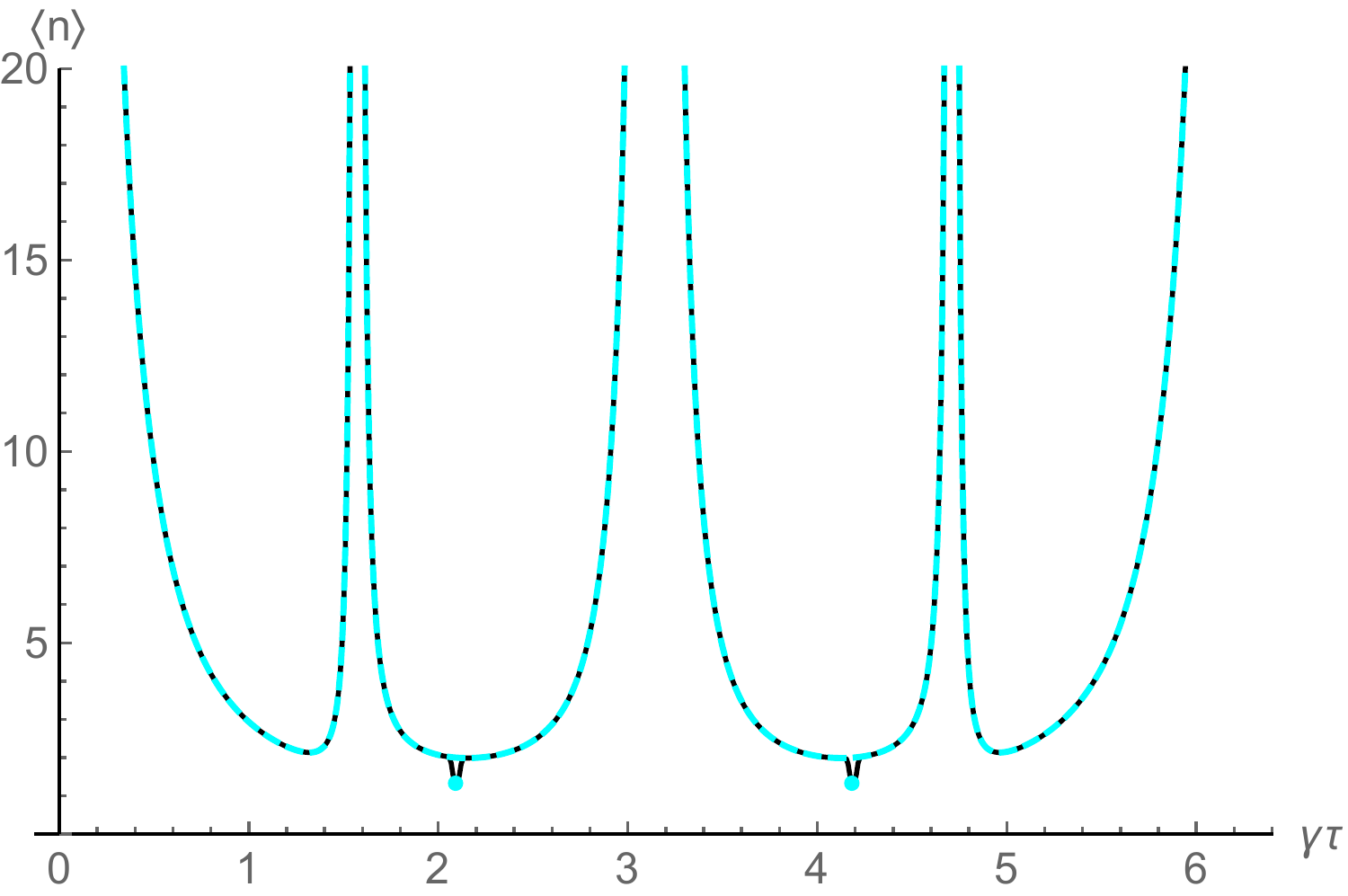}
\caption{
The average $\langle n \rangle$ versus $\gamma \tau$ Eq. (\ref{eqDAVID}).
 When
$\tau \to 0$ we find 
$\langle n \rangle \to \infty$ due to the Zeno effect, another
expected 
divergence of $\langle n \rangle$ is found when the sampling time is the full
 revival 
time $\gamma \tau =2 \pi$. In addition to these two points we find singularities
also for $\pi/2,\pi,3\pi/2$.
Notice the discontinuities of $\langle n \rangle$
 for  $\gamma \tau=2 \pi/3$ and
$\gamma \tau=4 \pi/3$  which are discussed in the text. 
Here $|\psi(0)\rangle=|3\rangle$ and the detection is on the origin.
The plot of $\langle n \rangle$  for this choice of initial condition
is  vastly different from that presented in
Fig.
\ref{fig2four}.
}
\label{fig2three}
\end{figure}

In contrast, if the  starting  state is $|3\rangle$
 the total probability of detection is found to be $1$, if the measurement
time $\tau$ is not the full revival time $\gamma \tau=2 \pi$,
 or  one of the exceptional
sampling times  listed in Table 
\ref{TableExcpetional}.
In Appendix \ref{AppKessler},
we find 
\begin{widetext}
\begin{equation}
\langle n \rangle =
{4 \over 9}+  { 9 \over 8 - 8  \cos\gamma \tau} + {1 \over 36 \cos^2 \gamma \tau} - {1 \over 9 \cos \gamma \tau} + {17 \over 72\left( 1 + \cos \gamma \tau \right)}
\label{eqDAVID}
\end{equation} 
\end{widetext}
an equation valid for all $\gamma \tau$ besides the exceptional points.
The general behavior of $\langle n \rangle$
is  obviously quite different from the case when the
initial location is $0$,  compare Fig.  
\ref{fig2four} and Fig. \ref{fig2three} indicating that the initial
condition plays a crucial rule.
As shown in Fig. 
\ref{fig2three} 
the average $\langle n \rangle$
exhibits nontrivial behavior as it diverges as it  approaches some of the
exceptional points. These singularities  are found near those
exceptional
sampling times where the total probability of measurement is not one.
Interestingly the values of $\langle n \rangle$, conditioned on return, are finite at the exceptional points
themselves.

An analytical calculation
for the exceptional sampling times $\gamma \tau = 2 \pi/3$ or $4 \pi/3$
 finds
$\langle n \rangle=4/3$. 
This sampling time is unique since the average $\langle n \rangle$ exhibits
a discontinuity: for $\gamma \tau$ in the vicinity of $2 \pi/3$ and $4 \pi/3$
 we
find using  
Eq. (\ref{eqDAVID})
$\langle n \rangle =2$ (so at these points the equation is not valid). 
Similar to any discontinuity at a point,
the discontinuity  of  $\langle n \rangle$ at $\gamma \tau=2 \pi/3,4 \pi/3$
 might not be detectable in experiment. However  one finds
critical slowing down, namely 
the  convergence of $\langle n \rangle$ for any point in the vicinity
of these exceptional points is very slow, 
as demonstrated in Fig. 
\ref{fig2three}.

\subsection{Rings of Size $L$}

 While the benzene ring is instructive, one must wonder how general
are the main results. In Appendix \ref{AppKessler}
 we derive the following 
four results:
\begin{itemize}
\item[i.] For a ring of size $L$ and for a particle initially on
site $x=0$ where the measurements are performed, the particle is
detected with probability  unity, and in this sense the motion is
recurrent. We emphasize that this result
is a property of the specified initial condition. 
\item[ii.] For the same initial condition, besides those isolated exceptional 
sampling times $\tau$ listed below, the average number of 
detection events is
\begin{equation}
\langle n \rangle =\left\{
\begin{array}{c c} 
{ L + 2 \over 2} & \  \mbox{$L$ is even} \\
\ & \ \\
{ L + 1 \over 2} & \  \mbox{$L$ is odd}. 
\end{array}
\right.
\label{eqKRSL}
\end{equation}
This result is remarkable since the average is independent of the sampling time $\tau$.
 Here we see that $\langle n \rangle$ is the number of distinct energy
levels of the system. 
In the language of \cite{Grunbaum} it is the winding 
number of the Schur function, or the effective dimension of the Hilbert space.
 For large systems $\langle n \rangle$ grows linearly with the size
of the system $L$, while from classical random walk
 theory we naively expect diffusive
scaling $\langle n \rangle \sim L^2$. In that sense the quantum
walk is more efficient (see below further remarks).
\item[iii.] The exceptional sampling times $\tau$ are given by the rule
\begin{equation}
\Delta E \tau = 2 \pi n
\label{eqET}
\end{equation}
where $n$ is an integer, and $\Delta E= E_i - E_j> 0$ is the energy
difference between pairs of eigenenergies of the underlying Hamiltonian
$H$. For example the stationary energies of the benzene ring 
are $\{ - 2 \gamma, - \gamma , \gamma , 2 \gamma\}$ as mentioned,
and hence Eq.  
(\ref{eqET}) predicts the exceptional sampling times
$0, \pi/2\gamma, 2 \pi/3\gamma , \pi/\gamma, ...$. 
The condition Eq. (\ref{eqET}) implies a partial revival of the wave packet
free of measurement, namely  two   modes of the system 
behave identically when strobed at the period $\tau$.
On these exceptional points the solution exhibits non-analytical behavior.
This is manifested in discontinuities or diverging behavior of $\langle n \rangle$
or the fluctuations of $n$ and also slow critical-like convergence
to the asymptotic theory.  The
exact nature of the non-analytical behavior depends on the initial
condition as we have demonstrated for the benzene ring. 

\item[iv]  For a particle starting on $|0\rangle$
every time the condition
(\ref{eqET}) is met by a pair of energy levels  
we reduce the value of
$\langle n \rangle$  by unity.
Thus Eq. (\ref{eqET}) is the upper limit of $\langle n \rangle$ for a system
of fixed size $L$. More specifically we find that 
\begin{equation}
\langle n \rangle = \mbox{number of distinct phases
} \exp( -i E_k \tau)
\label{eqPHASES}
\end{equation}
where $E_k$ are the energy levels of the system.
For nearly any $\tau$ this is the same as the number of distinct
 energy levels,
but of course for special sampling times, this integer is less
than that.

\end{itemize}

\subsection{Bose-Einstein Distribution}

A curiosity is the fact that we may express the solution for the $0$ (starting point) to $0$ (measurement point) problem for a general ring with $L$ sites,  in terms
of a Bose-Einstein distribution. The latter is defined
as \cite{Reif}
\begin{equation}
\bar{n}_k = {1 \over e^{-\beta \mu + \beta E_k} -1}
\label{eq32}
\end{equation}
where $\beta$ is the inverse temperature and $\mu$ the chemical potential.
In our case $z^{-1} =\exp(- \beta \mu)$ and $\beta E_k= i \tau E_k$
with the energy spectrum  $E_k = - 2 \gamma \cos( 2 \pi k/L)$. 
The generating function for a ring system with $L$ sites is 
\begin{equation}
\hat{\phi}(z) ={ \sum_{k=0} ^{L-1} \bar{n}_k/L \over 1 + \sum_{k=0} ^{L-1} \bar{n}_k /L }. 
\label{eq32a}
\end{equation}
The mathematical
 relation of the problem at hand to the Bose-Einstein distribution
is not limited to the specific energy spectrum under investigation
(see also \cite{Grunbaum}).
If the Hamiltonian is symmetric in the sense
that all lattice points
are equivalent, such that $|C_k|^2=1/L$, the above result is valid.
In the Bose-Einstein language
the sum $\sum_{k=0} ^{L-1} \bar{n}_k/L$, 
is the spatially averaged
density. Thus the problem
of finding the generating function is mathematically equivalent to finding
the relation between chemical potential, temperature and the 
average number
of particles, for a given energy spectrum of a system.
 The main conditions are that all sites
are equivalent and  
that the initial and detection state are both on a single ring site.

\subsection{Revivals}

 The amplitude of detection at the first measurement $\phi_1$ is now
investigated 
for a particle on a ring of size $L$ starting on the origin
which is also the detected site 
\begin{equation}
\phi_1 = {1 \over L} \sum_{k=0} ^{L-1} e^{2 i \gamma \tau \cos\left( 2 \pi k /L\right)}.
\label{eq32b}
\end{equation}
In the limit $L\to \infty$ the sum is an integral
and  we find $\phi_1=J_0 (2 \gamma \tau)$ and $J_0(\cdot)$ 
is a Bessel function of the first kind \cite{ABR}. 
  Unlike the
benzene,  $L=6$, case, for an infinite system
 with $\gamma \tau\neq 0$,
the probability of detecting the particle at
first measurement, $F_1=|\phi_1|^2$, is always less than unity 
since $|J_0(2 \gamma \tau)|^2<1$.
This is to be expected, as the wave function in an infinite 
system does not revive
at the origin. The question remains:
does a finite sized system always exhibit a special 
choice or choices  of $\gamma \tau \neq 0$ such that $F_1=1$?
(and  then
 all $F_n$ for $n>1$ are zero). This corresponds
to a deterministic outcome of certainly  detecting
 at a single detection attempt.
This question put differently is the well-studied
 question of full revivals.
Namely, does there exist some $\tau$ such that a particle, 
 in the absence
of measurement,  will fully return to its initial state.
If that is the case, 
the first measurement detects the particle with probability 
one if the measurement time is $\tau$. As mentioned, for $L=6$,
this effect is found  for $\gamma \tau = 2 \pi k$ for $k=0,1,2,\cdots$.

 According to \cite{Blumen} full revivals take place for $L=1,2,3,4,6$ only.
This can be verified using our formalism.
We checked this for $L=5,7,8,9,10$
finding that 
 the absolute value squared of the  sum 
Eq. (\ref{eq32b}) is never equal 1 unless $\tau=0$.
For example for $L=10$
\begin{equation}
\phi_1 = {1 \over 5} \left[ \cos \left( 2 \gamma \tau \right) +
4 \cos\left( {\gamma \tau\over 2} \right) \cos\left( \sqrt{5} \gamma \tau \over 2\right)\label{eq43aa}
\right],
\end{equation}
an expression which shows that $|\phi_1|< 1$ beyond the trivial
case $\gamma \tau=0$. 

{\bf Remark} 
For a ring with $4$ sites and sampling rate of $\gamma \tau=\pi$
one finds $\hat{\phi}(z)=z^2$  namely the quantum walker is
detected in the  second measurement with probability one.
Such a behavior is found when the  initial wave packet is localized
at the place of detection.
 In this example,  the first measurement is performed
when the wave function at the origin is zero, and
hence this measurement does not alter the wave function, while
in the second measurement we have full revival of the wave function
at the origin.

\section{First detection time for an unbounded quantum walker}
\label{secFD}

 In this section we consider the first detection problem for
a free particle in an infinite lattice. 
We use the  tight-binding Hamiltonian
\begin{equation}
H = - \gamma \sum_{x=-\infty}^{\infty} \left(|x\rangle \langle x+1 | + 
|x+1  \rangle \langle x| \right)
\label{eq01}
\end{equation}
  for a particle launched from the origin
$|\psi(0)\rangle = | 0 \rangle$ and
investigate the probability  of
first  detection $F_n$ with projective
measurements performed at the origin.

  Previously  Bach, et al. \cite{Bach}.
investigated a Hadamard quantum walk introduced in  \cite{Ambainis} 
showing that the survival probability of a one dimensional
walker  
exhibits a power-law decay, $F_n \propto n^{=3}$ in our terminology, namely
 a scaling
 exponent $3$ which is twice the classical 
one, i.e., $3/2$.
 This was the topic of further, analytical \cite{Grunbaum} numerical
\cite{Das,Miquel} and perturbative approaches \cite{Dhar,Dhar1}.
 In this sense, it is known already that  quantum walks modify the
known classical  exponents of classical random walk theory, where the first
passage time PDF decays like $t^{-3/2}$  for large times. 
What seems to be missed in previous literature, is that even in an infinite
system critical sampling effect takes place. On the more technical level we show how to use
the generating function formalism to find the large $n$ behavior of
the  first detection probability then finding non trivial behaviour of $S_\infty$.

\subsection{The generating function}

 The solution of the Schr\"odinger equation (\ref{eq01}) is 
$|\psi_f(t)\rangle= \sum_{x=-\infty} ^\infty B_x |x\rangle$ and
the amplitudes satisfy $i \dot{B}_x=- \gamma(B_{x+1} + B_{x-1})$.
Using the Bessel function identity \cite{ABR}
 $2 J'_\nu(z) = J_{\nu-1} (z) - J_{\nu+1}(z)$
and the initial condition $B_x= \delta_{x,0}$
one finds
\begin{equation}
|\psi_f(t)\rangle =
 \sum_{x=-\infty} ^\infty i^{x} J_x (2 \gamma t) | x \rangle.
\label{eq33}
\end{equation}
This is of course the same as 
 $|\psi_f(t)\rangle = \exp( - i H t) |0\rangle$. 
%
To obtain the generating function we use Eq. 
(\ref{eqRed06}), we have $\langle 0 |\hat{\psi}_f(z)\rangle= \sum_{n=0} ^\infty z^n J_0( 2 \gamma \tau n)$ so
\begin{equation}
\hat{\phi}(z) = 1 - {1 \over \langle 0 |\hat{\psi}_f(z)\rangle} =
1 - {1 \over \sum_{n=0} ^\infty z^n J_0 ( 2 \gamma \tau n)}.
\label{eq35Zan}
\end{equation}
Since $J_0(0)=1$ this can be rewritten as
\begin{equation}
\hat{\phi}(z) = { \sum_{n=1} ^\infty z^n J_0 \left( 2 \gamma n \tau\right) \over  
1 +  \sum_{n=1} ^\infty z^n J_0 \left( 2 \gamma n \tau\right) }.
\label{eq35}
\end{equation}

Before analyzing Eq. (\ref{eq35}), we 
derive it again  using the energy
spectrum.
 The energy levels of a tight-binding  ring system 
(periodic boundary conditions) 
determined from the time independent Schr\"odinger equation,  are
$E_k = - 2 \gamma \cos( 2 \pi k/L)$ and the
system size $L$ tends to infinity.
As we have seen,  $|C_k|^2 = 1/L$ in Eq. (\ref{eq30}),
since all lattice sites are equivalent with respect to the Hamiltonian. 
Using Eq. (\ref{eq22}) or Eq. (\ref{eq32a})  with minor rearrangement
\begin{equation}
\hat{\phi}(z) = { 
{ 1 \over L} \sum_{k=0} ^{L-1} { z \exp\left[ i 2 \gamma \tau \cos\left( { 2 \pi k \over L} \right)\right] \over 1 - z \exp\left[ i 2 \gamma \tau \cos\left(   2 \pi k \over L \right) \right]}
\over 
1 + 
 { 1 \over L} \sum_{k=0} ^{L-1} { z \exp\left[ i 2 \gamma \tau \cos\left( { 2 \pi k \over L} \right)\right] \over 1 - z \exp\left[  i 2 \gamma \tau \cos\left(   2 \pi k \over L \right) \right]}
}.
\label{eq36}
\end{equation}
This is exact for all $L$  and reduces to Eq. (\ref{eq31}) when $L=6$. 
Let 
\begin{equation}
I(z)\equiv 
 \lim_{L \to \infty}
{ 1 \over L} \sum_{k=0} ^{L-1} { z \exp\left[ i 2 \gamma \tau \cos\left( { 2 \pi k \over L} \right)\right] \over 1 - z \exp\left[  i 2 \gamma \tau \cos\left(   2 \pi k \over L \right) \right]
}.
\label{eq37}
\end{equation}
where the sum is an integral in the limit.  Changing variables
$2 \pi n/L=y$ 
\begin{equation}
I(z) = {1\over  2 \pi} \int_0 ^{2 \pi} 
{ 
z e^{ 2 i \gamma \tau \cos(y) } {\rm d} y \over 
1 -  z e^{ 2 i \gamma \tau \cos(y) } 
}
\label{eq38}
\end{equation}
and expanding to get a geometric series
gives
\begin{equation}
I(z) = { z \over 2 \pi }
 \int_0 ^{2 \pi} e^{  2 i \gamma \tau \cos(y)} \sum_{k=0} ^\infty \left[ z e^{ 2 i \gamma \tau \cos\left( y \right)} \right]^k {\rm d} y. 
\label{eq38A}
\end{equation}
Integrating over $y$
using the identity 
$\int_0 ^{2 \pi} \exp[ i z \cos(y)] {\rm d} y = 2 \pi J_0 (z)$
and shifting the summation by unity, we get 
\begin{equation}
I(z) = \sum_{n=1} ^\infty z^n J_0 \left( 2 \gamma n \tau \right).
\label{eq39}
\end{equation}
Using 
\begin{equation}
\hat{\phi}(z) = {I(z)\over 1 + I(z)} 
\label{eq40}
\end{equation}
we find  the generating function Eq. 
(\ref{eq35}). Note that $1+ I(z) = \langle 0|\hat{\psi}_f(z)\rangle$, and we use
it as a matter of convenience.

\subsection{Small $n$ behaviour}

 To analyze the small $n$ behavior of $\phi_n$, we expand the generating
function 
as a power series of $z$ using
Eqs. (\ref{eq18},\ref{eq35}).
Such an expansion is easy to perform with a symbolic program like Mathematica,
which provides Table \ref{Table01}  giving explicit expressions
for  $\phi_n$ when  $n=1,\cdots,7$.
If we set $\gamma \tau$ to a fixed value, the expansion can be carried out
for relatively large $n$, and in this sense we may find  
numerically exact results which are later presented in the figures.  
Of course this information is the same as that found with the
exact solution Eqs.  
(\ref{eq16},\ref{eq17}).

From Table \ref{Table01}, we see that 
 when $\gamma \tau \to 0$ we have $\phi_n=\delta_{n1}$ since then
the particle is detected in the first attempt.
The Table also gives
the leading order corrections to this expected behavior
\begin{equation}
\phi_n \sim \left\{
\begin{array}{c l}
 1 - (\gamma \tau)^2 &  n = 1 \\
  - 2 (\gamma \tau)^2 &  n \geq 2.
\end{array}
\right.
\label{eqDEM}
\end{equation}
This can 
be derived
from Eq. (\ref{eq35}) using $J_0(x) \simeq 1 - x^2/4$ for $x\ll 1$.
Eq. (\ref{eqDEM}) exhibits equal probability of detection
for $n>1$ which is merely an outcome of the Taylor expansion.

When
$\gamma \tau \to \infty$ 
the amplitudes
$\phi_n$ are zero for $n=1,2,3,\cdots$
since the wave packet spreads in an infinite system,
hence its probability of being detected on the origin is zero in this limit.
A close look at Table \ref{Table01} shows that
in this limit $\phi_n \simeq J_0 \left( 2 n \gamma \tau\right)$.
Using $F_n = |\phi_n|^2$ and the asymptotic behavior
of the Bessel function 
\begin{equation}
F_n \sim 
{ \cos^2\left( 2 \gamma \tau n - \pi/4\right)
\over \pi \gamma \tau n}
\label{eqPPHH}
\end{equation}
when $\gamma \tau \gg 1$.
Thus the probability of first detection decays like a $1/n$ power law,
when $\gamma \tau$ is large and $n$ is fixed and finite.
This approximation breaks down for sufficiently large $n$. For example,
using $\gamma \tau =80$,
by comparison with the numerically exact solution, we observe roughly
$30 \%$ deviation from theory already for $n=10$

\begin{widetext}

\begin{table}[h!]
\centering
\begin{tabular}{| >{$}c<{$} | >{$}l<{$}|}
\hline
n &\phi_n\\
\hline
1 &J_0(2 \gamma \tau )\\
\hline
2 &J_0(4 \gamma \tau )-J_0(2 \gamma \tau ){}^2\\
\hline
3 & J_0(2 \gamma \tau ){}^3-2 J_0(4 \gamma \tau ) J_0(2 \gamma \tau )+J_0(6\gamma \tau )\\
\hline
4 &-J_0(2 \gamma \tau ){}^4+3 J_0(4 \gamma \tau ) J_0(2 \gamma \tau ){}^2-2 J_0(6 \gamma \tau ) J_0(2 \gamma \tau )-J_0(4 \gamma\tau ){}^2+J_0(8 \gamma \tau )\\
\hline
5 &J_0(2 \gamma \tau ){}^5-4 J_0(4 \gamma \tau ) J_0(2 \gamma \tau ){}^3+3 J_0(6 \gamma \tau ) J_0(2 \gamma \tau){}^2+\\
  &+3 J_0(4 \gamma \tau ){}^2 J_0(2 \gamma \tau )-2 J_0(8 \gamma \tau ) J_0(2 \gamma \tau )-2 J_0(4 \gamma \tau ) J_0(6 \gamma \tau)+J_0(10 \gamma \tau )\\
\hline
6 &-J_0(2 \gamma \tau ){}^6+5 J_0(4 \gamma \tau ) J_0(2 \gamma \tau ){}^4-4 J_0(6 \gamma \tau ) J_0(2 \gamma \tau){}^3-6 J_0(4 \gamma \tau ){}^2 J_0(2 \gamma \tau ){}^2+3 J_0(8 \gamma \tau ) J_0(2 \gamma \tau ){}^2+\\
 &+6 J_0(4 \gamma \tau ) J_0(6 \gamma\tau ) J_0(2 \gamma \tau )-2 J_0(10 \gamma \tau ) J_0(2 \gamma \tau )+J_0(4 \gamma \tau ){}^3-J_0(6 \gamma \tau ){}^2-2 J_0(4 \gamma \tau )
J_0(8 \gamma \tau )+J_0(12 \gamma \tau )\\
\hline
7 &J_0(2 \gamma \tau ){}^7-6 J_0(4 \gamma \tau ) J_0(2 \gamma \tau ){}^5+5 J_0(6 \gamma \tau ) J_0(2\gamma \tau ){}^4+10 J_0(4 \gamma \tau ){}^2 J_0(2 \gamma \tau ){}^3-4 J_0(8 \gamma \tau ) J_0(2 \gamma \tau ){}^3-\\
 &-12 J_0(4 \gamma \tau ) J_0(6 \gamma \tau ) J_0(2 \gamma \tau ){}^2+3J_0(10 \gamma \tau ) J_0(2 \gamma \tau ){}^2-4 J_0(4 \gamma \tau ){}^3 J_0(2 \gamma \tau )+3J_0(6 \gamma \tau ){}^2 J_0(2 \gamma \tau )+\\
  &+6 J_0(4 \gamma \tau ) J_0(8 \gamma \tau ) J_0(2 \gamma \tau )-2 J_0(12 \gamma \tau ) J_0(2\gamma \tau )+3 J_0(4 \gamma \tau ){}^2 J_0(6 \gamma \tau )-2 J_0(6 \gamma \tau ) J_0(8 \gamma \tau )-\\
  &-2 J_0(4 \gamma \tau ) J_0(10 \gamma\tau )+J_0(14 \gamma \tau )\\
\hline
\end{tabular}
\caption{Amplitudes  of $\phi_n$ for an infinite system, the quantum walker
dispatched and detected at the origin.}
\label{Table01}
\end{table}

\begin{figure}
\centering
\includegraphics[width=0.40\textwidth]{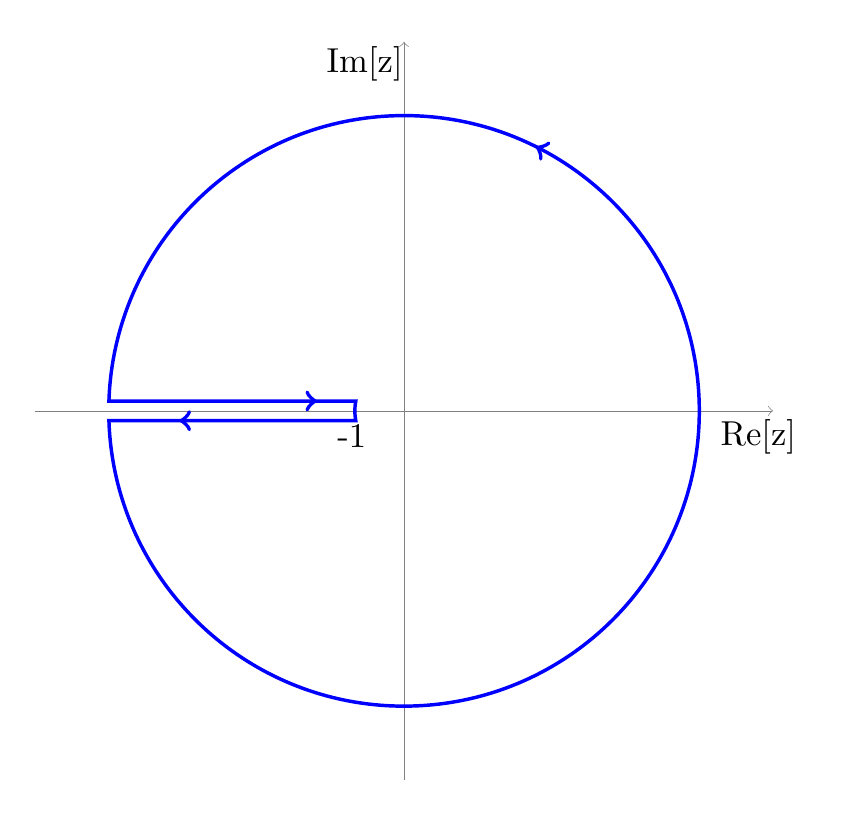}
\caption{
Counter-clockwise integration path $C$ in Eq. (\ref{eq23})
 for evaluation of $\phi_n$ for an unbounded lattice with the sampling rate
$\gamma \tau = \pi/2$  Eq. 
(\ref{eq46a}) avoids  the
branch cut along the negative real axis
when $|z|>1$. The outer radius approaches infinity.
}
\label{fig2path}
\end{figure}

\begin{figure}
\centering
\includegraphics[width=0.40\textwidth]{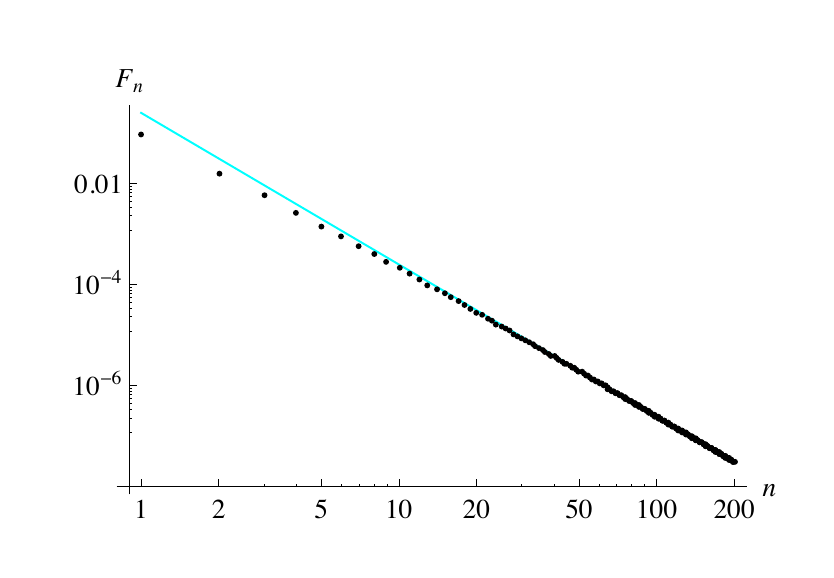}
\caption{
First detection probability $F_n$  versus $n$ on a log log plot,
for an open system, detection is at the starting point and
the sampling rate is $\gamma \tau=\pi/2$. For large $n$ the exact result (dots)
 converges to the
asymptotic power law behavior  Eq. 
(\ref{eq53}), $F_n =|\phi_n|^2 \sim 0.25 n^{-3}$ (straight line). 
}
\label{figsingle}
\end{figure}

\end{widetext}

\subsection{Large $n$ behaviour}

 The large $n$ behaviour of $\phi_n$  is of particular
theoretical interest  since it is expected to exhibit universal features.
For ordinary random walks the transformation from $z$ back to $n$
 is performed
with machinery called the Tauberian and Abelian theorems \cite{Redner,Weiss}.
While the technique is widely applicable 
the transformation method for the quantum problem 
is slightly more involved as compared 
with the corresponding  classical  problems, the
reason being that $\phi_n$ turns out to exhibit 
a decay superimposed with oscillations, while in classical random walks
the first passage probability decays monotonically to zero for large $n$. 

The analysis is performed by integration in the complex plane
using Eqs. (\ref{eq23},\ref{eq39},\ref{eq40}). In the large $n$ limit
the asymptotic behaviour of the Bessel function 
\cite{ABR}
is
\begin{equation}
J_0 \left( 2 \gamma \tau n \right) \sim { \cos\left( 2 \gamma \tau n - \pi/4\right) \over \sqrt{ \pi \gamma \tau n } }.
\label{eq42}
\end{equation} 
Since we are investigating the large $n$ limit we replace
the Bessel function 
in Eq. (\ref{eq39}) 
with its asymptotic 
behavior, hence we
define 
\begin{equation}
I_{\gamma \tau} (z) = \sum_{n=1} ^\infty z^n 
 { \cos\left( 2 \gamma \tau n - \pi/4\right) \over \sqrt{ \pi \gamma \tau n }}
.
\label{eq43}
\end{equation} 
This expression works well for $|z| \simeq 1$ corresponding to large $n$
(only $|z|>1$ is actually important, see our integration path below). 
The  large $n$ limit of $\phi_n$ is then given by the asymptotics
of the inverse $Z$-transform of 
\begin{equation}
\hat{\phi}(z) \sim { I_{\gamma \tau} (z) \over 1 + I_{\gamma \tau}(z) }.
\label{eq44}
\end{equation} 
We start with an example.

\begin{widetext}

\subsubsection{ Infinite system $\gamma \tau = \pi/2$} 

 We consider the case $\gamma \tau = \pi/2$ and find
\begin{equation}
I_{\pi/2} (z) = { 1\over \pi} \mbox{Li}_{1/2}(-z),
\label{eq45}
\end{equation}
where $\mbox{Li}_{s}(z) =\sum_{k=1} ^\infty z^k/k^s$ is the polylogarithm 
function.
Using Eq. (\ref{eq23})
\begin{equation}
\phi_n \sim {1 \over 2 \pi i} \oint_C z^{-n -1} 
 {  {1 \over \pi}\mbox{Li}_{1/2}(-z)  \over 1 + {1 \over \pi} \mbox{Li}_{1/2}(-z) } {\rm d} z
\label{eq46a}
\end{equation}
in the large $n$ limit.
The integration path is shown in Fig. \ref{fig2path}. 
A branch-cut is found in the complex plane of integration along the negative
real axis when 
$|z|>1$  since there $z=-|x|$ and $\mbox{Li}_{1/2}(-z)=\sum_{k=1} ^\infty |x|^k /\sqrt{k}$ does not converge. 
The radius of the outer
 path of integration is taken to be large 
($r \to \infty$ in Fig. \ref{fig2path})
and then 
\begin{equation}
\phi_n \sim  {{\cal I}_{+} + {\cal I}_{-} \over 2 \pi i } .
\label{eq46}
\end{equation}
The integration in the complex
plane reduces to two integrals running parallel to the
branch cut (see Fig.
\ref{fig2path}).
The first line integral ${\cal I}_{+}$
 to be evaluated is slightly above the negative
real axis  along  $z=x + i \epsilon$ with
$-\infty < x < -1$ and $0<\epsilon \to 0$.
The second integral follows in the opposite direction with $z=x-i \epsilon$
(see Fig. \ref{fig2path}).  We consider ${\cal I}_{+}$ using 
$z^{-(n+1)} = \exp[-(n+1) \ln z]$
and Eq. (\ref{eq46a}):
\begin{equation}
{\cal I}_{+} = \int_{-\infty} ^{-1} \exp\left[ - \left( 1 + n \right) \ln \left(  x + i \epsilon \right) \right] 
{ \mbox{Li}_{1/2} ( - x - i \epsilon) \over \pi + \mbox{Li}_{1/2} (-x-i \epsilon) }{\rm d} x. 
\label{eq47}
\end{equation}
${\cal I}_{-}$ is similarly defined with a change of sign in $\epsilon$ and
the lower and upper integration limits switched.
Changing variables to $y$ with $x\equiv -1 -y$ 
\begin{equation}
{\cal I}_{+} = \int_0 ^\infty \exp\left[ - \left(1 + n\right) 
\ln\left(-1 - y + i \epsilon\right)\right]
{ \mbox{Li}_{1/2} (1 + y - i \epsilon) \over 
\pi + \mbox{Li}_{1/2} (1 + y - i \epsilon) } {\rm d} y.
\label{eq48}
\end{equation}
When $n \to \infty$ clearly the small $y$ limit of the 
integration dominates.
Close to the singularity at $z=1$ \cite{MathematicaREF}
\begin{equation}
\mbox{Li}_{1/2}(z) \simeq \sqrt{ { \pi \over 1- z} } + \zeta(1/2) + \cdots,
\label{eq48zeta}
\end{equation}
where $\zeta(.)$ is the Riemann zeta function. Indeed
to obtain the leading term (which will eventually give the large $n$ limit of $\phi_n$)
  we replace the summation with integration in the definition of 
the Polylog function $\mbox{Li}_{1/2}(z)$  using
\begin{equation}
\sum_{k=1} ^\infty { z^k \over  \sqrt{k}} \simeq \int_0 ^\infty {{\rm d} k \over \sqrt{k} }  e^{ k \ln (z) } = \sqrt{\pi}\left[-\ln(z)\right]^{-1/2}  \simeq \sqrt{ \pi \over 1 - z },
\label{eq48ll}
\end{equation} 
where $z<1$. 
We use $z=1 + y \pm i \epsilon$, where the choice of sign depends on the path evaluated, namely ${\cal I}_{\pm}$. In the limit of small $y$, corresponding to large $n$, 
we find, using Eq. (\ref{eq48zeta}),
\begin{equation} 
\mbox{Li}_{1/2} \left(1 + y \pm i \epsilon\right) \simeq
\sqrt{{ \pi \over - y \mp i \epsilon}}\sim \sqrt{{\pi \over y e^{\mp i \pi}}}=
 \pm i \sqrt{{\pi \over y}}.
\label{eq49}
\end{equation}
Given that $\ln( - 1 - y +i \epsilon) = \ln(1+y) + i \pi$ when $\epsilon \to 0$, 
\begin{equation}
{\cal I}_{+} = \int_0 ^\infty \exp\left\{ - (1 + n) \left[\ln(1 + y) + i \pi\right]\right\} 
{ -i \sqrt{\pi/y } \over  
\pi  -i \sqrt{\pi/y } }
{\rm d} y . 
\label{eq50}
\end{equation}
Clearly $\exp[ - (1 +n ) i \pi] = (-1)^{n+1}$, and  approximating
 $(1+n)\ln(1+y) \sim n y$ 
in the exponential in the integrand in Eq. (\ref{eq50}),
an approximation  
 valid in the limit of large $n$ since then only small $y$ contributes
 to the integration, and finally Taylor 
expanding 
$ -i \sqrt{\pi/y } / [
\pi  -i \sqrt{\pi/y } ] \sim 1 - i \sqrt{\pi y}$ we find
\begin{equation}
{\cal I}_{+} \sim (-1)^{n+1} 
\int_0 ^\infty \exp( - ny) \left( 1 - i \sqrt{ \pi y} + \cdots\right) {\rm d} y.
\label{eq51}
\end{equation}
The integral yields
\begin{equation}
{\cal I}_{+}  \sim (-1)^{n + 1} \left( {1 \over n} - i {\pi \over 2 n^{3/2} } \right).  
\label{eq52}
\end{equation}
The calculation of ${\cal I}_{-}$ follows the same  steps 
\begin{equation}
{\cal I}_{-} \sim (-1)^{n+1} \left(-{1\over n}  - i {\pi \over 2  n^{3/2}} \right).
\label{eq53a}
\end{equation}
Finally,
using Eq. (\ref{eq46})
\begin{equation}
\phi_n \sim{ (-1)^n \over 2 n^{3/2} }.
\label{eq53}
\end{equation}
This solution exhibits odd/even oscillations with an overall decay
of a power law. The probability of finding the particle
after $n$ attempts goes like $F_n\sim 4^{-1} n^{-3}$. Hence
it does not exhibit oscillations, but that is merely due to our
choice of sampling rate $\gamma \tau=\pi/2$ as we now show.
In Fig. \ref{figsingle}, a very nice agreement between 
Eq. (\ref{eq53}) and the exact solution is seen 
already for not too large
$n$.

\begin{figure}
\centering
\includegraphics[width=0.39\textwidth]{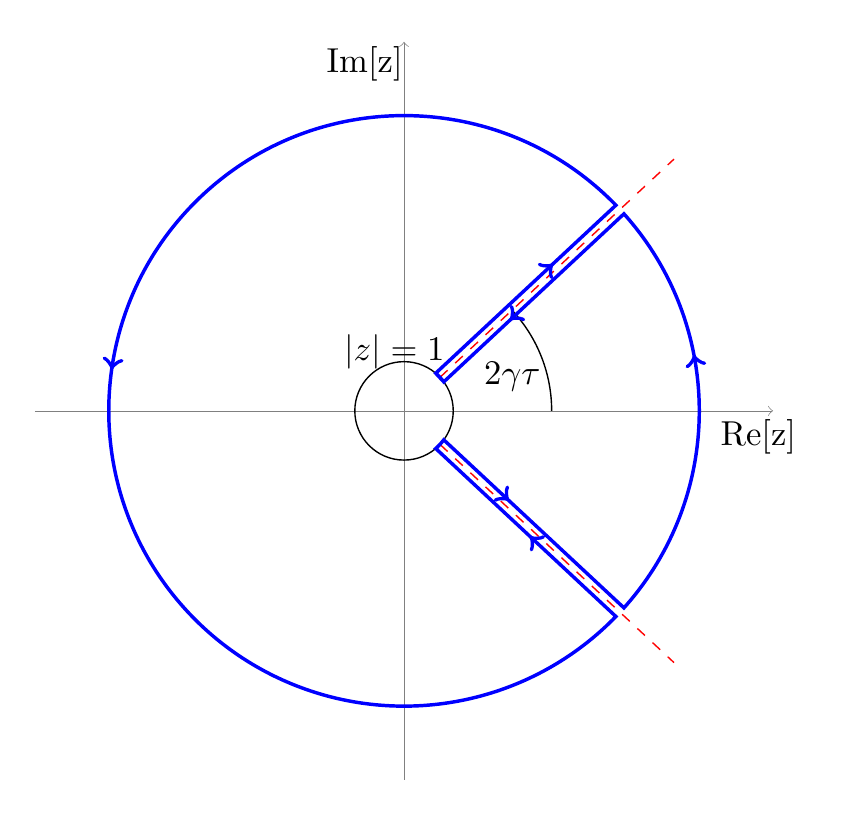}
\caption{
Integration path for the calculation of $\phi_n$
in the complex plane
bypasses  two branch cuts (dashed lines).
Integration along four lines just above and below the two dashed lines
is explained in the text, while the integration around the outer circle does not
contribute in the limit of an  infinite radius.
 When $2 \gamma \tau= k \pi$ the two branch cuts merge $(k=0,1,\cdots)$. 
}
\label{fig1path}
\end{figure}

\begin{figure}
\centering
\includegraphics[width=0.49\textwidth]{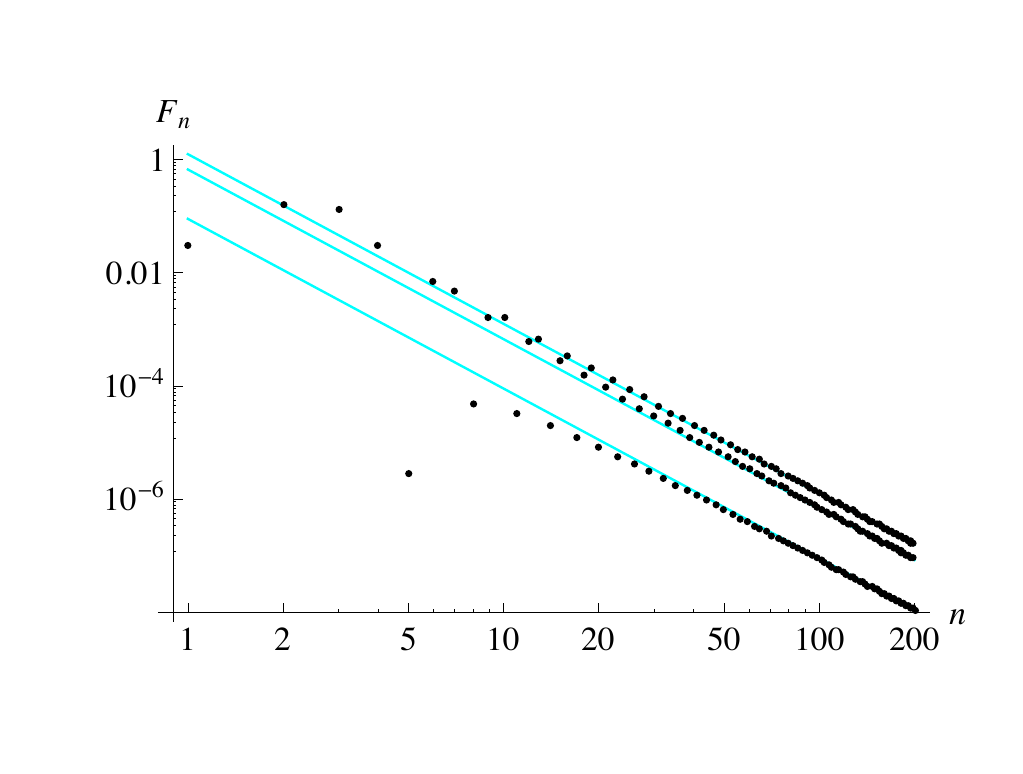}
\caption{
$F_n$ versus $n$ for the rational  sampling time  
$\gamma \tau /\pi=1/3$. Now the
detection probability $F_n$ 
 decays monotonically like $n^{-3}$ with a superimposed
periodic oscillation.  The lines are the asymptotic theory 
Eq. 
(\ref{eqGen11})
 which for large $n$ 
nicely match the exact expression (dots). 
}
\label{fig1patha}
\end{figure}

\subsubsection{First detection statistics in $1d$ for an open system} 

 We now investigate  $\phi_n$  for sampling time
 $0<\gamma \tau<\pi$ with $\gamma \tau\neq \pi/2$,
sticking to the case where
the origin of the quantum walk is also the location where the 
particle is detected.  We find the large $n$ limit of
$\phi_n$  using
\begin{equation}
\phi_n \sim {1 \over 2 \pi i} 
\oint_C z^{-n -1} {I_{\gamma \tau} (z) \over 1 + I_{\gamma \tau} (z) } {\rm d} z
\label{eqGen01}
\end{equation}
and similar to the previous sub-section the large argument limit of the Bessel
function 
Eq. (\ref{eq42})
gives
\begin{equation}
I_{\gamma \tau} (z) = \sum_{n=1} ^\infty z^n { \cos\left( 2 \gamma \tau n - \pi/4\right) \over \sqrt{ \pi \gamma \tau n}} =
{1 \over 2 \sqrt{2 \gamma \tau}}
 \left[
e^{- i \pi/4} \sum_{n=1} {\left( z e^{2 i \gamma \tau}\right)^n \over \sqrt{n}} +
e^{ i \pi /4} \sum_{n=1} ^\infty  {\left( z e^{ - 2 i \gamma \tau}\right)^n \over \sqrt{n}} \right].
\label{eqGen02}
\end{equation}
Clearly for $z=r \exp(i \theta)$ with $\theta=2 \gamma \tau$ or $\theta=- 2 \gamma \tau$ and $r\geq 1$
either the second or first sums diverge respectively.
 Thus for $0<2 \gamma \tau<2 \pi$ we find two branch cuts
which are  shown in Fig. \ref{fig1path}. The exception is the case
treated in the previous sub-section  $2 \gamma \tau=\pi$
where the two branch cuts merge. 
The integration path in the
 complex plane now avoids two branch cuts, but otherwise
the calculation is similar to the one we performed in the  previous section. 
In Appendix \ref{AppSolve}
we find one of our main results
\begin{equation}
\phi_n \sim 2 \sqrt{ {\gamma \tau \over \pi n^3}} \cos\left( 2 \gamma \tau n + {\pi \over 4} \right) . 
\label{eqGen10}
\end{equation} 
Thus the probability of measuring the quantum walker returning to its
origin for the first time is 
\begin{equation}
F_n \sim { 4 \gamma \tau \over \pi n^3} \cos^2 \left( 2 \gamma \tau n + {\pi \over 4} \right).
\label{eqGen11}
\end{equation} 
This formula predicts that when $\gamma \tau/\pi$ is a rational number the
probability $F_n$ multiplied by $n^3$ is periodic. Such a behavior
is shown in Fig. \ref{fig1patha}.
In contrast, if $\gamma \tau/\pi$ is not rational, the asymptotic behavior
is quasiperiodic and appears noisy, the theory Eq. (\ref{eqGen11})
 perfectly matching the exact solution. This we demonstrate in Figs.
\ref{figNum}
for the  choice of  irrational $\gamma \tau /\pi=0.8/\pi$. 

A strange aspect of Eq. (\ref{eqGen11}) 
is that in the limit of $2 \gamma\tau \rightarrow
\pi$ it
does not recover the result found for $2 \gamma\tau = \pi$
$F_n \sim 0.25 n^{-3}$ found in Eq. 
(\ref{eq53}). Instead $\lim_{2 \gamma \tau \to \pi} F_n\sim n^{-3}$
so a factor four mismatch is found.  
 This surprising result
is no doubt due to the presence of two branch cuts for the case under study, 
while for $2 \gamma \tau=\pi$ we have only one. 
This implies that the convergence to formula Eq. (\ref{eqGen11})
when $2 \gamma \tau \simeq \pi$  is very slow, and is reminiscent of the
behavior at the exceptional sampling time we saw for finite $L$. 

 These calculations can be extended for $\gamma \tau> 2 \pi$
 and critical behavior is found
 for $\gamma \tau=k \pi/2$ with $k=1,2,\cdots$,
 since then the two branch cuts merge, and
one finds $F_n \sim k /(4 n^3)$. Otherwise  Eq. 
(\ref{eqGen11}) is valid for all sampling periods $0<\gamma \tau$. 

 Note that the energies on a finite  tight-binding  ring are given by 
$E_k=-2 \gamma \cos(2 \pi k /L)$, as mentioned. Hence in the limit 
of large $L$ we
find  a band of energies
of size $\Delta E=4 \gamma$. If we use this width in  Eq.   
(\ref{eqET}) we find the exceptional sampling time $2 \gamma \tau= k\pi$.
This argument leads us to speculate that the width of the band, in an infinite system,
will determine the exceptional points that survive the $L\to \infty$ limit.

\end{widetext}

\begin{figure}
\centering
\includegraphics[width=0.49\textwidth]{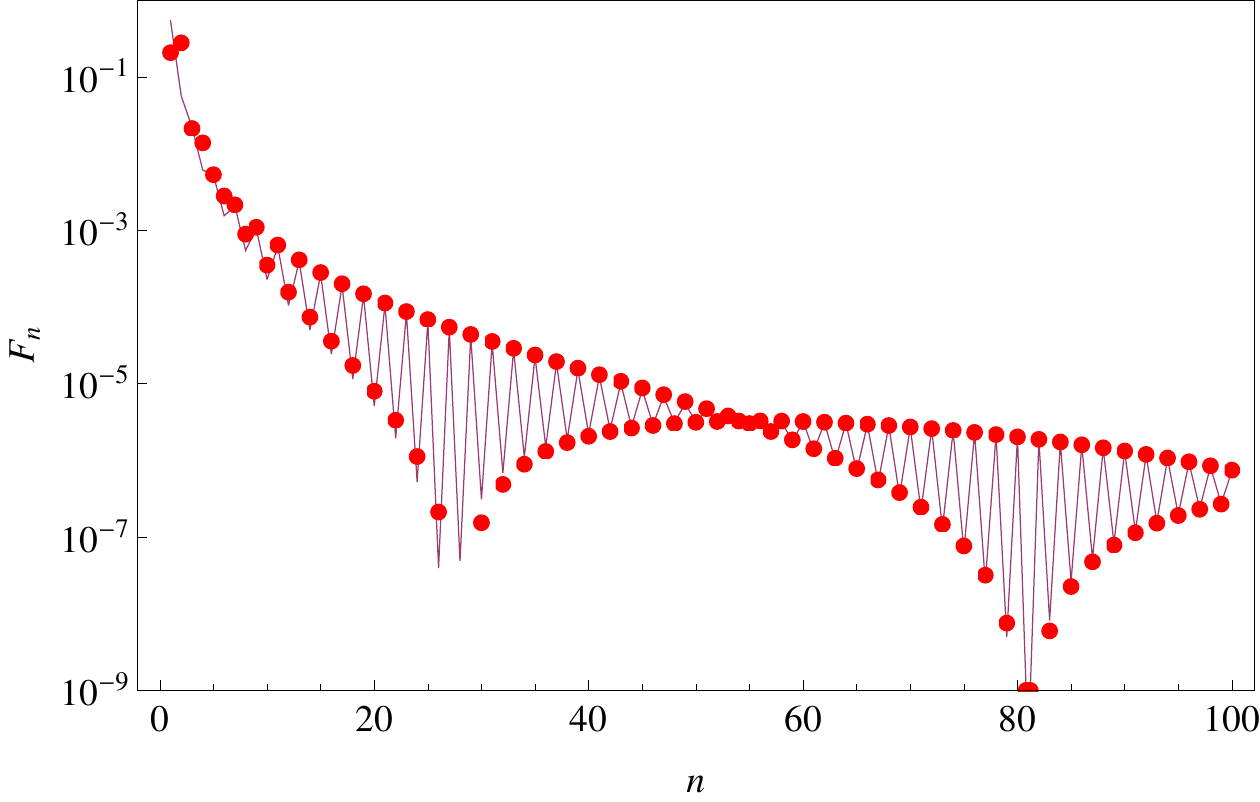}
\caption{
$F_n$  versus $n$ for the irrational choice of sampling rate $\gamma \tau/\pi =0.8/\pi$, for 
a quantum walk on a one dimensional lattice (note the log linear scale).
The numerically exact solution (red circles)  nicely
matches the theory (the curve) already for moderate values
of $n$.  
}
\label{figNum}
\end{figure}


\subsection{$1d$ Quantum walks are not recurrent, the survival
probability is highly irregular}
 
 The probability that the quantum walker
 will eventually be detected on its
origin is given by 
$1-S_\infty=\sum_{n=1} ^\infty F_n$. Here 
$S_\infty$ is the probability that the particle survived, 
namely the probability that it was not detected.  
For a one dimensional classical
 random walk on the integers, i.e. the binomial random walk, where the
 particle has probability $1/2$ to jump left or right, 
the survival probability is zero, so eventually the
 particle is detected at its origin. 
 The quantum walk in one dimension is generally non-recurrent
as previously pointed out \cite{Bach,Ambainis}.
The spreading is ballistic, not diffusive, and hence the return to the origin
is not guaranteed in an open system. 

 We focus therefore on the non-trivial value of the survival
probability  $S_\infty$. 
Using the  exact expressions for $F_n$ we have used Mathematica to obtain
estimation for $1-S_\infty$ using two methods.  The first is summing $F_n$
for a large value of $n$ using the exact expression for $F_n$.
More precisely, we expand
the generating function $\hat{\phi}(z)$
 in $z$ the coefficients giving $\phi_n$
up to some large vale of $n$, and hence also $F_n$.  
Then we estimate the reminder using 
our asymptotic large $n$ formulas. 
The second method we numerically perform the integration in Eq. 
(\ref{eqTot}), using Eq. (\ref{eq35}).
  Both
methods yield the same results.

\begin{figure}
\centering
\includegraphics[width=0.49\textwidth]{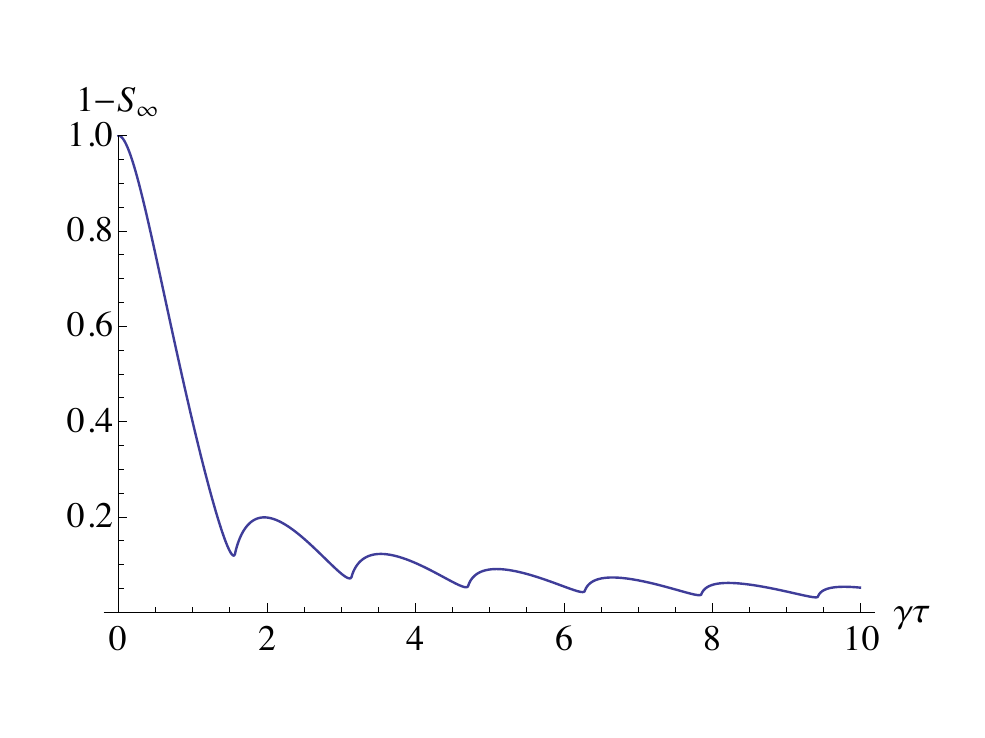}
\caption{
For the one dimensional tight-binding quantum walk the
 probability that the particle
is eventually detected $1-S_\infty=\sum_{n=1} ^\infty F_n$ 
versus the sampling rate $\gamma \tau$ exhibits a non-monotonic behavior. 
Unlike its classical random walk
counterpart, the quantum  walk is not recurrent,
 unless $\gamma \tau \to 0$, which is the
trivial case.
}
\label{figFGInf}
\end{figure}

 In Fig. 
\ref{figFGInf} we show $1 - S_\infty$ versus $\gamma \tau$.
 For $\gamma \tau \to 0$ we get $S_\infty=0$ since 
the particle starting at the origin is detected with 
probability one if the measurement is made immediately
 after the release of the particle.
 Not surprisingly, when $\gamma \tau \to \infty$ $S_\infty\to 0$ (though it remains small
though finite for $\gamma \tau \simeq 10$). An unforeseen property is 
the cusps in 
 $1-S_\infty$, presented in the figure, 
  found for $\gamma \tau = \pi k/2$ with $k=1, 2, \cdots$.
Mathematically these cusps must be related to the appearance of two branch
 cuts in the complex plane.
 The non-monotonic behavior of the probability of eventual measurement
is not something we
could have anticipated.

  For large $\gamma \tau$ we find  $F_n \propto 1/n$ 
for finite $n$ and $F_n \propto n^{-3}$ for large $n$,
Eqs. 
(\ref{eqPPHH})
and 
(\ref{eqGen11}). Matching these solutions we expect a transition to
found when  $n_{tr}= c \left(\gamma \tau\right)$ where $c$ is a constant
of order unity.
We may estimate the probability of detection as  $\sum_{n=1} ^{n_{tr}} F_n$.
Here we do not sum in the interval $(n_{tr},\infty)$ since here
$F_n$ decays like 
$n^{-3}$ and hence negligible. Using Eq. 
(\ref{eqPPHH}), switching from summation to integration,
we find 
\begin{equation}
1 - S_\infty \simeq {\ln \gamma \tau \over 2 \pi \gamma \tau}
\label{eqSUMINT}
\end{equation}
when $\gamma \tau\gg 1$. In this rough estimation we used
$\ln \gamma \tau \gg  \ln c$. Further discussion on $S_\infty$ and a far better approximation
is provided
in Appendix D.

\subsection{Transition to large system}

\begin{figure}
\centering
\includegraphics[width=0.49\textwidth]{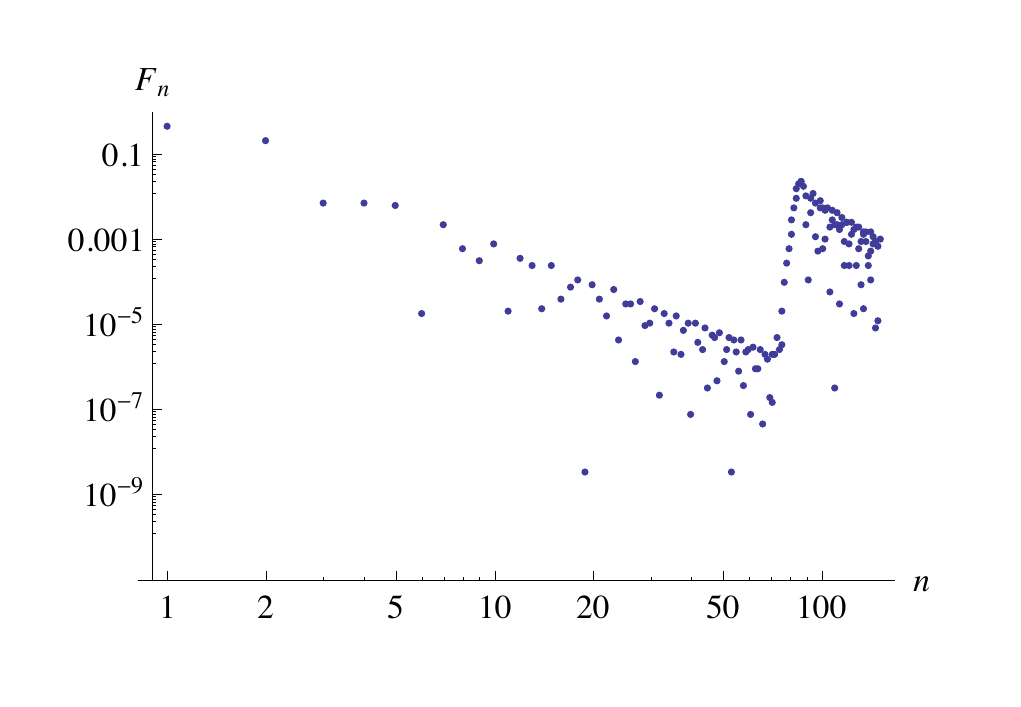}
\caption{
Quantum walk on a ring with 
 $L=100$ sites for a sampling rate $\gamma \tau=0.6$ with detection at the
origin of the walk. 
For small $n$ the system exhibits a behavior similar to that of an infinite system
(see Fig. \ref{fig4ex}). Roughly at $n=70$, there is a
sudden increase in $F_n$, 
probably due to a partial revival of the packet at the origin. 
}
\label{fig4}
\end{figure}

\begin{figure}
\centering
\includegraphics[width=0.49\textwidth]{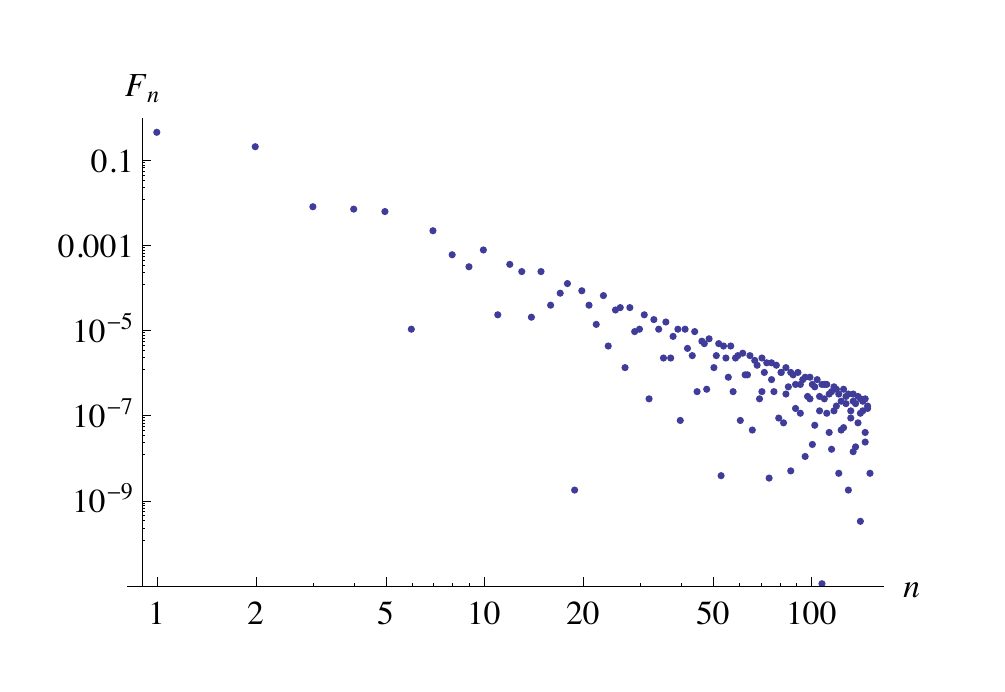}
\caption{
Same as Fig. \ref{fig4} for an infinite system. 
}
\label{fig4ex}
\end{figure}

 For a large but finite ring of size $L$ we expect to see a
behavior similar 
 to the infinite system, for intermediate times, where on the one
hand the asymptotic limit Eq. 
(\ref{eqGen11})
is reached, but the particle still does
not sense the finiteness of the system. This behavior is presented
in Fig. 
\ref{fig4} for a ring with $L=100$ sites. At first $F_n$ decays
as for the infinite system with the same sampling rate (see
 Fig. \ref{fig4ex}). 
However, at least in this example, roughly at $n \simeq 70$ we see a sudden
increase in $F_n$. This is a non-classical behavior; 
for a classical random walk on a large ring the survival probability has a power-law decay for intermediate times, crossing over to an exponential decay for long times.
 The significant increase in $F_n$ is due to a partial revival 
at the origin, a non-classical effect, and is obviously related to the ballistic
nature of the quantum walk.
The precise nature of this transition merits further study.

\section{Other approaches and some connections}
\label{secOTHER}

%
 Krovi and Brun \cite{Brun06,Brun07} derived a
general expression for the average hitting
time $\langle n \rangle$  for discrete quantum walks. 
Their formalism is based on a trace formula for a density matrix, while we
relate the statistics of the first detection event to the
wave function free of measurements. They also point out to the possibility
of an  infinite average hitting times on finite graphs. This theme
is important in the context of the efficiency of search. A classical
random walker, on a finite graph, always finds the target (assuming the walk
is ergodic). In that sense a classical walk is very efficient since
it will always reach its target. On the other hand, a  classical walk
tends to explore territory previously visited, so the time it takes
to reach the target is relatively long. Quantum walks
are considered faster, if compared to classical walks, in the
sense that they scale ballistically. However, as we showed 
for the simple benzene ring geometry, 
the average hitting time (for specific initial conditions) may diverge.
It implies that
one cannot categorically say that quantum search, in the average
hitting time sense, is more efficient than the classical counterpart.
However,  for the ring geometry, 
and for initial condition on the origin, Eq.
(\ref{eqKRSL})
 shows that the average $\langle n \rangle$  scales with the size of system,
which is a ballistic feature of the search, not a diffusive one. 
It should be noted that diverging $\langle n \rangle$ for finite systems are found here for the stroboscopic measurement under investigation. The latter has 
many advantages, for example in revealing quantum periodicities,
 but if the goal is to detect the particle with probability one,
 measurements
on times drawn from a Poisson process should be more efficient \cite{Brun08}.

As mentioned in the introduction the quantum first passage time question is controversial
in the sense that one has many approaches to the problem.
One way to treat the quantum first passage time problem, 
is by adding a non-Hermitian term to the Hamiltonian \cite{Krap}.
 Briefly the non-Hermitian approach leads
to the non-conservation of the  normalization of the wave function, 
which can be interpreted 
as the survival probability of the particle.
This  approach  was shown to be related to the projective measurement 
method in the limit of small finite $\tau$ \cite{Dhar,Dhar1}. 
It is an interesting question however whether
the non-Hermitian approach can predict the behaviors found in this manuscript,
for example Eq. (\ref{eqGen11}).
Especially important is whether 
the limits of large $n$  and small $\tau$ commute, an issue
left   
to future research. 

Dhar et al. \cite{Dhar,Dhar1}, after laying the
 general framework to the problem,
including introducing the fundamental concept of the wave function $\phi_n$, 
 investigated 
the limit of small $\tau$. In this sense
the problem is investigated close to the 
Zeno limit. The stroboscopic sampling for finite
$\tau$  investigated herein allows for revivals, critical sampling, and other special quantum 
effects which are  clearly missed in this Zeno limit.
They \cite{Dhar,Dhar1} use a perturbation theory
and  derive an effective Hamiltonian
where the effects of the failed prior detections 
are  included as a non-Hermitian part of the Hamiltonian. 
 That method together with numerical simulations
 predicts, for a specific initial
condition and a finite sized system,  that the survival probability has a power law decay of 
$-1/2$ or $-3/2$. This is an interesting observation since it
shows that in some cases
 half integer exponents control the decay of the survival probability, while so far we have found only integer values.
 We have verified 
 results obtained by Dhar et al. using the
generating function formalism with exact numerical method (not shown).  
Oscillations or anomalous measurement periods are not apparent under the restriction of small $\tau$.
It might be possible,
to prove analytically, 
that in the small $\tau$ limit, one detects transients different from
our prediction for $F_n$
Eqs. 
(\ref{eq53},
\ref{eqGen10}). As mentioned, the oscillating power-law
 tails of $F_n$  vanish when $\tau\to 0$,
and hence correction terms become very important when $\tau$
is small. In other words, perhaps there are two universality classes
of exponents for quantum random walks. The current picture is that we have
four exponents $1$, $3$, $1/2$ and $3/2$ which depend on the initial
position, the value of $\tau$,  and the geometry.

Another approach suggests the use of the classical renewal equation, derived by
Schr\"odinger \cite{Schro} long before the appearance of wave mechanics in 
the context of the first passage time of a Brownian motion, to investigate 
also the
 quantum first passage time problem
 \cite{Lumpkin,Ranjith}. 
This approach seems very different than ours, 
since it does not take into consideration the effect of measurement, 
and  it uses probabilities instead of amplitudes to 
describe the quantum statistical aspects of the first passage  problem.

 In \cite{Stefanak} a discarding system  method
was used 
for  the problem of recurrence of quantum walks. 
These authors investigated 
 an ensemble of identically prepared systems and suggested
the following measurement procedure. After one time step measure the occupancy
of the particle at the origin (the outcome is a binary yes/no answer)
 and then discard the system. 
Take a second identically prepared second system
and let it evolve for two time steps, measure at the origin and again
discard the system. Continue similarly for a long time. 
One then constructs the probability of measurement at the origin,
for times $t=1,2,...$. In this way one may define a P$\hat{\mbox{o}}$lya number,
which in classical random walks gives the criterion whether
a random walk
is transient or recurrent. This approach is of course very different than ours.
 As pointed out in Ref.  \cite{Stefanak}, 
one can imagine several approaches to the problem of recurrence
of a quantum random walker. 

{\bf Historical remark.}
Eq. (10) in Ref.  \cite{Muga} is the renewal equation for classical
 random walks after which  the author writes:
``The equation was proposed by Schr\"odinger in terms of cumulative probabilities
."
A minor historical remark is that
a  close look at the original publication
 \cite{Schro} does not reveal an explicit equation.
However, translating the original work to English reveals that indeed
the origin of the renewal equation is in that classic paper. 
In Appendix B of \cite{Muga}
Muga and Leavens mention
 a quantum renewal equation, their conclusion is that
this equation (which is not at all related to ours)
is not valid.

\section{Summary }

The quantum first detected passage time problem 
 rests on two fundamental
postulates of quantum reality. The first is the 
 Schr\"odinger equation and the second is the projective measurement
postulate. The latter is the fifth postulate listed in \cite{CT}
which states that immediately after the measurement the state of the
system is a normalized projection of $|\psi\rangle$, where
$|\psi\rangle$ is the state function of the system just prior to the measurement.
Without these assumptions, the first tested in many experiments but
the second in far fewer,  quantum theory is not complete.
Our goal, following \cite{Dhar}  was not to question basic physical
 assumptions,
but rather show how these postulates lead to the solution of the
first detection  problem.  
 
The quantum renewal equation (\ref{eq15}) and the $Z$ transforms
Eqs. (\ref{eqRed06},
\ref{eqRed06aa})
give
a rather general relation between the amplitude of first detection $\phi_n$
and the wave function of the system free of measurement $|\psi_f\rangle$.
In that sense the problem of first detection
reduces to the solution of the
 Schr\"odinger 
  equation,
e.g. the determination of the energy spectrum. In particular,
similar to the corresponding classical first passage  problems,
the generating function is a powerful
 tool with which we can attain many insights. 

We have illustrated several surprising features, both for closed
systems, like rings and for open systems. Generally the quantum first detection problem exhibits behaviors very different than classical, but still some
relations remain. For example for a classical random walk on an infinite
line,  the first passage 
PDF decays like $n^{-3/2}$, similarly the corresponding 
quantum amplitude
(neglecting the oscillations) gives  $\phi_n \sim n^{-3/2}$ [see
Eqs. 
(\ref{eq53},
\ref{eqGen10})].
However the quantum problem exhibits rich behaviors, which are related to the sampling rate $\gamma \tau$,
including oscillations of $F_n$ superimposed on the power law decay, 
the Zeno effect when $\tau \to 0$ and a surprising critical behavior when 
$\gamma \tau=k\pi/2$. Thus the
first detection problem is critically sensitive to the sampling rate even 
for an infinite
system and even in the large $n$ limit. Of course for real systems 
this conclusion can be reached only if the coherence is 
maintained for long times. 
Another notable result is the non-analytical behavior
of the survival probability $S_\infty$ 
 as we tune $\gamma \tau$, 
see Fig. \ref{figFGInf}.
This should be compared 
with the classical result where the final probability of detecting
the particle in one dimension is unity. 

For a finite system like a ring, we find strong sensitivity to
the initial condition: for an initial
condition starting on the origin $x=0$,
which is also the detected site, 
the  average number of detection attempts $\langle n \rangle$, 
increases  linearly  with the system size $L$, see Eq.
(\ref{eqKRSL}).
Further $\langle n \rangle$ does not depend
 on the sampling time $\tau$, besides ever present exceptional
sampling,
see Eq. 
(\ref{eqET}).
 Since $\langle n \rangle \propto L$ we find
ballistic scaling of $\langle n \rangle$ compared
with diffusive scaling for the corresponding classical problem. 
 However,  when starting on other initial conditions, $\langle n \rangle$
 depends crucially
on the sampling time $\tau$, in fact  $\langle n \rangle$ may
diverge as $\tau$ is tuned. 
The revivals, optimal detection times, exceptional points given
by the simple formula Eq.  
(\ref{eqET}),
half
dark states,  and quantization of $\langle n \rangle$ for the
$|0\rangle\rightarrow |0\rangle$ transition \cite{Grunbaum},
 describe rich behaviors even in small systems.
They also point out the advantage of stroboscopic observation
of the system, since this captures underlying quantum features. 
Even for a finite sized system,
 the probability of being eventually detected is not
always unity. However, we showed that at least when starting on the
origin which is also the location of the detection, the particle
is detected with probability unity for any sampling time $\tau$, 
and in that sense the quantum
walk on a finite ring or more generally a graph \cite{Grunbaum}
is recurrent. 
Finally, the tools developed in this
paper can serve as starting point for many other first detection problems
and the advance of single particle first detection theory. 

{\bf Acknowledgement}  We thank the Israel Science Foundation, grant number 376/12, for funding and Janos Asboth for pointing out Ref. \cite{Grunbaum}.
EB thanks Abhishek Dhar for discussions and hospitality at ICTS Campus
in Bangalore.

\newpage

%
%


\newpage

\appendix
\begin{widetext}
\section{Iterations, general measurements}
\label{AppITGM}

In this Appendix, we present the details of the derivation of our general formulation.
\subsection{Derivation of $\phi_n$, $\hat{\phi}(z)$}
We here show by induction that 
\begin{equation}
\label{eq:direct_formula}
\ket{\theta_n}=[U(\tau)(1-\hat{D})]^{n-1}U(\tau)\ket{\psi(0)}
\end{equation}
is identical to 
\begin{equation}
\label{eq:general_recursion}
\ket{\theta_n}=U(n\tau)\ket{\psi(0)}-\sum_{k=1}^{n-1}U[(n-k)\tau]\hat{D}\ket{\theta_k}.
\end{equation}
The case $n=1$ can be easily verified  since both equations give
$\ket{\theta_1}= U(\tau)\ket{\psi(0)}$.
We now argue by induction,
assuming  the validity of the Eq. (\ref{eq:general_recursion})  for some 
$n$, and  proving it  for $n+1$.
 As can be seen from Eq. (\ref{eq:direct_formula}), the effective 
waveform is propagated between measurements 
\begin{align}
\ket{\theta_{n+1}}&=U(\tau)(1-\hat{D})\ket{\theta_n}\nonumber\\
&=
U(\tau)U(n\tau)\ket{\psi(0)}-\sum_{k=1}^{n-1}U[(n+1-k)\tau]\hat{D}\ket{\theta_k}-U(\tau)\hat{D}\ket{\theta_n}\nonumber\\
&=
U\big((n+1)\tau\big)\ket{\psi(0)}-\sum_{k=1}^{n+1-1}U\big((n+1-k)\tau\big)\hat{D}\ket{\theta_k}.
\end{align}
This is of course the same 
as Eq. (\ref{eq:general_recursion}) but for $n+1$ hence the proof is completed.

The $Z$ transform of Eq. (\ref{eq:general_recursion}), gives
 a closed formula for
$\ket{\theta(z)}=\sum_{n=1}^{\infty}z^n\ket{\theta_n}$. 
Using the convolution theorem, we find
\begin{equation}
\ket{\theta(z)}=[1+U(z)\hat{D}]^{-1}U(z)\ket{\psi(0)}.
\label{eq:general_z_transform}
\end{equation}
%
For a single detected site at $0$ ($\hat{D}=\ket{0}\bra{0}$) and
 denoting $\phi_n=\bracket{0|\theta_n}$, Eqs.
(\ref{eq:general_recursion}, 
\ref{eq:general_z_transform}) can be represented as:
\begin{equation}
\phi_n=\bra{0}U(n\tau)\ket{\psi(0)}-\sum_{k=1}^{n-1}\bra{0}U[(n-k)\tau]\ket{0}\phi_k,   \ \ 
 \ \ \  \hat{\phi}(z)=\frac{\bra{0}U(z)\ket{\psi(0)}}{1+\bra{0}U(z)\ket{0}},
\end{equation}
respectively, as it is stated in the text.

\end{widetext}

\subsection{General measurements}

 We consider the first detection  measurement of an observable whose
corresponding bra is 
$\langle {\cal O} |$ with the additional condition
$\langle{\cal O}|{\cal O}\rangle =1$ so
$\sum_{x \in X} \langle x|$
is not such a measurement if the set $X$ has more then one element,
 but $\langle x|$  or
$\langle E_m|$, denoting an energy state of the system, are.
For example 
we select a specific energy level denoted $E_m$
 and ask what is the statistics of first detection
time of that state,
so here $\langle {\cal O} | = \langle E_m |$ (the subscript $m$ is for measurement). 
 The energy states are assumed to be non-degenerate for simplicity. 
The generating function in this case is 
\begin{equation}
\hat{\phi}(z)  = {\langle {\cal O} | \hat{U}(z) | \mbox{initial} \rangle \over 1 +
\langle {\cal O} | \hat{U}(z) | {\cal O} \rangle}.
\label{eq24}
\end{equation}
Here the state $ \langle \cal{O}|$ is not only normalized but it must
be an eigenstate of an Hermitian operator, in such a way that it
describes a physical measurement. 
For example consider the case where the initial state is a stationary state of the Hamiltonian
$|E_i\rangle$ and the observable state is $|{\cal O}\rangle = |E_m \rangle$.
 Then 
the assumption that the Hamiltonian is time independent
means that
\begin{equation}
\langle E_m | \hat{U}(z) | E_i \rangle
 = \left[ z^{-1} \exp\left(  i E_m \tau\right) -1\right]^{-1} \delta_{mi}
\label{eq25App}
\end{equation}
where $\delta_{mi}$ is the delta of Kronecker, we find
\begin{equation}
\hat{\phi}(z) = z e^{-i E_m \tau} \delta_{mi}.
\label{eq26}
\end{equation}
This is the expected result, if we start with a stationary state 
$i$ this state will be detected with probability one only if
$m=i$. More than one measurement is actually not informative in this
case, hence $\phi_n=0$ for $n>1$ (this is easily understood since the generating function contains a single term linear in $z$ when $m=i$). 

It is emphasized again 
that the  measurement is different from the standard textbook measurement
that asks what is the energy of the system at a certain time (say $\tau$).
In our case the measurement performed gives a binary answer either yes or no,
and this gives a definitive answer to the question whether the system is   in 
the $m$-th state at times $\tau,2 \tau .... $. 

\begin{widetext}
\section{Analytic Calculations of Moments of $F_n$ For a $L$-site Ring}
\label{AppKessler}

\subsection{$0 \to 0$}
We start with the problem where the particle is both released from and detected at the origin.  There is a simple proof that ${\cal{F}} \equiv \sum_n F_n = 1-S_\infty = 1$.  
Eq. (\ref{eq32a}) gives
\begin{equation}
\hat{\phi}(z) = \frac{\sum_{k=0}^{L-1}\bar{n}_k}{L + \sum_{k=0}^{L-1} \bar{n}_k}.
\label{eqA1}
\end{equation}
Now, for $z=e^{i\theta}$
\begin{equation}
\bar{n}_k = \frac{1}{e^{i(\tau E_k-\theta)}-1}=\frac{\cos(\tau E_k-\theta) -1-i\sin(\tau E_k-\theta)}{2-2\cos(\tau E_k-\theta)} =
-\frac{1}{2} - \frac{i}{2} \cot \frac{\tau E_k - \theta}{2}.
\end{equation}
Thus, on the unit circle, $z=e^{i\theta}$, $\hat{\phi}(z)$ has the form
\begin{equation}
\hat{\phi}(e^{i \theta}) = \frac{-L/2 + iA}{L/2 + iA}
\end{equation}
where $A$ is real.   Thus $|\hat{\phi}(e^{i \theta})|^2 = 1$, and so, using Eq. (\ref{eqTot}),
\begin{equation}
{\cal{F}} = \sum_n F_n = \int_0^{2\pi} \frac{d\theta}{2\pi} |\hat{\phi}(e^{i \theta})|^2 = 1.
\end{equation}
and the particle is detected with probability unity.

For $z$ off the unit circle, $\hat{\phi}(z)$ can be written as a rational function of $z$, that is to say the quotient of two polynomials.  In fact, we have
\begin{equation}
\hat{\phi}(z) = \frac{  {\cal{N}}(z)}{{\cal{D}}(z)} .
\end{equation}
The exact form of the numerator and denominator depend on whether $L$ 
is even or odd, since in the former case we have $K=L/2+1$ different $E_j$'s, of which all but two are doubly degenerate, whereas in the latter case we have $K=(L+1)/2$ $E_j$'s, all but one of which are doubly degenerate.  We treat the $L$ odd case in detail, the even case being similar.  Note that we do not need in the following the specific values of the energies
 for the different states, so we absorb $\tau$ into the energies for efficiency.  In particular,
using Eq. (\ref{eqA1})
\begin{align}
{\cal{N}}(z)&=\left[\prod_{i=0}^{K-1} (e^{i E_i} - z)\right] \sum_{j=0}^{K-1}  \frac{d_j z}{e^{i E_j}-z} ;\nonumber\\
{\cal{D}}(z)&=\left[\prod_{i=0}^{K-1} (e^{i E_i} - z)\right] \left[L + \sum_{j=0}^{K-1} \frac{d_j z}{e^{i E_j}-z}\right].
\label{ND}
\end{align}
Here $d_j$ is the degeneracy of the $j$-th energy level, so that $d_0=1$ and otherwise $d_j=2$.
The denominator is a polynomial of degree $K-1$ since the $z^{K}$ term cancels out, while the numerator is of degree $K$, and has no $z^0$ term.  For the moment, we treat ${\cal{N}}(z)$ and ${\cal{D}}(z)$ as two order $K$ polynomials in $z$ defined by Eq. \ref{ND}. The two polynomials both have complex coefficients, each depending on the same set of real numbers $\{E_0,\dots,E_{K-1}\}$. What we want to show is that the two polynomials are related, for arbitrary $K$, by the relation
\begin{equation}
{\cal{D}}_K(z) = (-1)^{K-1}e^{i \sum_j E_j} z^{K} {\cal{N}}_K^*(1/z) .
\label{NDrelate}
\end{equation}
where we have made explicit the dependence of the polynomials on $K$, and have left implicit the dependence on the set of numbers $\{E_j\}$, $j=0,\ldots,K-1$.  The polynomial ${\cal{N}}^*$ is the polynomial with coefficients conjugate to those of ${\cal{N}}$.  Equivalently, this is the polynomial with all the $\{E_j\}$ replaced by $\{-E_j\}$.
For example, for $K=1$,
\begin{equation}
{\cal{N}}_1(z) = z; \qquad\qquad {\cal{D}}_1(z) =  (e^{i E_0} - z) + z = e^{i E_0},
\end{equation}
which clearly obey the relation Eq. (\ref{NDrelate}), for arbitrary real $E_0$.  We will now prove Eq. (\ref{ND}) by induction.  We start by noting the most of the terms in ${\cal{N}}_K$ and ${\cal{D}}_K$ also appear in ${\cal{N}}_{K-1}$, ${\cal{D}}_{K-1}$ with the set of energies $\{E_j\}$, $j=0,\ldots,K-2$. We
have
\begin{align}
{\cal{N}}_K(z) &= \left(e^{iE_{K-1}}-z\right) {\cal{N}}_{K-1}(z) + 2 z\prod_{i=0}^{K-2} (e^{i E_i} - z);\nonumber\\
{\cal{D}}_K(z) &= \left(e^{iE_{K-1}}-z\right) {\cal{D}}_{K-1}(z) + 2\prod_{j=0}^{K-1} \left(e^{iE_j}-z\right)  +  2 z\prod_{i=0}^{K-2} (e^{i E_i} - z) \nonumber\\
&= \left(e^{iE_{K-1}}-z\right) {\cal{D}}_{K-1}(z) + 2e^{iE_{K-1}}\prod_{j=0}^{K-2}\left(e^{iE_j}-z\right)   .
\label{A8}
\end{align}
We can now use the induction hypothesis, assuming Eq. (\ref{NDrelate}) is valid for $K-1$, to rewrite the first line of Eq.( \ref{A8}):
\begin{align}
(-1)^{K-1} z^K e^{i\sum_j E_j}{\cal{N}}_K^*(1/z) &= (-1)^{K-1}z^Ke^{i\sum_j E_j}  \left[\left(e^{-iE_{K-1}}-1/z\right)
{\cal{N}}_{K-1}^*(z) + (2/z)\prod_{i=0}^{K-2} (e^{-i E_i} - 1/z) \right]\nonumber\\
&= -ze^{iE_{K-1}} \left(e^{-iE_{K-1}}-1/z\right){\cal{D}}_{K-1}(z) + 2ze^{iE_{K-1}} \prod_{i=0}^{K-2} (e^{i E_i} - z)\nonumber\\
&=  \left(e^{iE_{K-1}}-z\right){\cal{D}}_{K-1}(z) + 2e^{iE_{K-1}} \prod_{i=0}^{K-2} (e^{i E_i} - z)\nonumber\\
&={\cal{D}}_K(z).
\end{align}
where we have invoked the last line of Eq. (\ref{A8}) in the final step.
This completes the induction proof.

As an immediate corollary, we obtain
\begin{equation}
\hat{\phi}^*(1/z) \hat{\phi}(z) = \frac{{\cal{N}}^*(1/z)}{{\cal{D}}^*(1/z)} \times \frac{{\cal{N}}(z)}{{\cal{D}}(z)} = 1,
\end{equation}
which implies $|\hat{\phi}(e^{i \theta})|^2=1$, which we previously obtained.

An alternative route
to prove Eq. 
(\ref{NDrelate})
 is to use Eq. (\ref{ND}) and write
\begin{equation}
{\cal N}(z) = \sum_{j=0} ^{K-1} d_j z \mathop{\prod_{i=0} ^{K-1}}_{i\ne j} \left( e^{i E_i} - z \right)
\end{equation}
and so
\begin{align}
{\cal D}(z) &= L   \prod_{i=0} ^{K-1} \left( e^{ i E_i} -z \right) + {\cal N} (z) \nonumber\\
 &= L \prod_{i=0} ^{K-1} \left( e^{ i E_i} -z\right) +
\sum_{j=0} ^{K-1} d_j \left[ \left( z - e^{i E_j}\right) + e^{i E_j}\right]
\mathop{\prod_{i=0}^{K-1}}_{i\ne j} \left( e^{i E_i} -z\right).
\end{align} 
Clearly
\begin{equation}
\sum_{j=0} ^{K-1} d_j \left( z - e^{ i E_j} \right) \mathop{\prod_{i=0} ^{K-1}}_{i\neq j}
 \left( e^{ i E_i} - z \right) =
- \left( \sum_{j=0} ^{K-1} d_j \right) \prod_{i=0} ^{K-1} \left( e^{i E_i} - z\right) ,
\end{equation}
and using $\sum_{j=0} ^{K-1} d_j = L $ we find
\begin{equation}
{\cal D} (z) = \sum_{j=0} ^{K-1} d_j e^{i E_j} \mathop{\prod_{i=0}^{K-1}}_{i\neq j} \left( e^{i E_i} - z\right).
\end{equation}
Thus ${\cal D}(z)$ is the same as ${\cal N}(z)$ when the replacement
$d_j z \rightarrow d_j \exp(i E_j)$ is made. It is now easy to verify
the theorem Eq.  
(\ref{NDrelate}). 

This factorization of $\hat{\phi}(z)$ allows for a simple calculation of $\langle n\rangle$, the mean detection time. The one added piece of information we require is the location of the zeros of ${\cal{D}}$.  It is clear that all 
these zeros lie outside the unit circle, as otherwise, $S_\infty$ would diverge.  For large $n$, $\phi_n$ decays geometrically as $r^{-n}$, where $r$ is the absolute value of the radius of the pole nearest to the origin. For the sum of $|\phi_n|^2$ to converge, we must have $r>1$.  There is one exception to this rule.  It turns out that for a discrete set of exceptional values of $\gamma \tau$, one (or in the case of even $L$, a complex conjugate pair) zero hits the unit circle.  Given the relationship between the numerator ${\cal{N}}$ and the denominator ${\cal{D}}$, a zero of ${\cal{N}}$ must hit the unit circle and and coincide with the zero of ${\cal{D}}$ at the exceptional point.  In this case, there is no pole in $\hat{\phi}(z)$ at this point, and all poles of $\hat{\phi}(z)$ still lie strictly outside the unit circle. We will return to the identification of these exceptional values of $\gamma\tau$ in a moment, but let us first proceed and calculate $\langle n \rangle$ for a non exceptional
$\gamma\tau$.
Given our theorem Eq. (\ref{NDrelate})  relating ${\cal{N}}$ and ${\cal{D}}$, we have
\begin{equation}
\hat{\phi}(z) = ze^{-i{\sum_j E_j}} \prod_{i=0}^{K-1} \frac{z-1/z^*_i}{(z-z_i)/z_i},
\end{equation}
where, as before $K=(L+1)/2$ for $L$ odd and $L/2+1$ for $K$ even, and the $z_i$ are the zeros of ${\cal{D}}(z)$.  Then, 
\begin{equation}
\hat{\phi}^*(1/z)\frac{d}{dz} \hat{\phi}(z) = \frac{1}{z} + \sum_{i=1}^{K-1} \left[\frac{1}{z-1/z_i^*} - \frac{1}{z-z_i}\right].
\end{equation}
We have to integrate this over the unit circle, which by the residue theorem picks up a contribution of $2\pi i$ for each pole in the interior, which lie at $0$ and $1/z_i^*$. Thus, as long as we are not dealing with an exceptional point, we have
\begin{equation}
\langle n\rangle = K.
\end{equation}
This agrees with explicit numerical calculations for $L=5, 6$.  If the exceptional point is such that a single pole touches the unit circle (a real pole for even $L$ or a complex one for odd $L$), then $\langle n\rangle$ is reduced by one.  If the exceptional point is such that a complex conjugate pair touch the unit circle, $\langle n\rangle$ is reduced by two at this value of $\gamma\tau$.

\subsection{Exceptional $\tau$}
For $z$ on the unit circle, we may write $z=e^{i\theta}$.  Studying the structure of ${\cal{D}}$, it is clear that
one way to make ${\cal{D}}$ vanish is to require that two of the factors $e^{iE_k}-z$ are zero, in which case every individual term in ${\cal{D}}$ vanishes separately.  In other words, we have, for a pair of energies $E_j < E_k$, 
\begin{equation}
\theta = E_j\tau + 2\pi n_j = E_k\tau + 2\pi n_k,
\end{equation}
for two integers $n_j$, $n_k$.  This gives us
\begin{equation}
\tau  = 2\pi n_{jk}/(E_k-E_j); \qquad\qquad \theta = \mod(E_j \tau, 2\pi)
\end{equation}
for an integer $n_{jk}$.
Thus for each pair $j,k$, there is an infinite number
  of exceptional values of $\tau$. For $L=6$, for example,  we have $\langle n \rangle =4$ for non-exceptional points.  As the energy levels in this case are $\{-2\gamma,-\gamma,\gamma,2\gamma\}$, we have an exceptional $\tau=\theta=0$, with degeneracy 3, so $\langle n\rangle$ is reduced by 3.  A second exceptional value is $\gamma\tau=\pi/2$, $\theta=\pi$, which comes from the pair $\{E_0, E_3\}$ with degeneracy 1. We also have $\gamma\tau=\pi$, $\theta=\pi$, coming both from the pair $\{E_0, E_3\}$ with $n_{jk}=2$ and also from $\{E_1, E_2\}$ so that this root has degeneracy 2. In addition, we have
$\gamma\tau=2\pi/3$, with $\theta=2\pi/3, 4\pi/3$, coming from $\{E_0,E_2\}$ and $\{E_1, E_3\}$ respectively, so that $\langle n\rangle$ is again reduced by 2.

\subsection{$L=6$, $3 \to 0$}
Unfortunately, there does not appear to be any such miracle occurring for the $L=6$ ring when the particle starts at $3$ and we detect at $0$,   the $3 \to 0$  transition.  Again, $\hat{\phi}(z)$ can be written as a rational polynomial, but we have not found any simple relationship between the numerator and denominator.  Nevertheless, we can still compute the moments of $n$.  We start by factoring ${\cal{D}}$,
\begin{equation}
\hat{\phi}(z) = i \frac{z\hat{\cal{N}}(z)}{{\cal{D}}(z)} = -i \frac{z\hat{\cal{N}}(z)}{D_0(z-z_1)(z-z_2)(z-z_3)},
\end{equation}
where we have also factored out $z$ and a phase from the numerator,  such that $\hat{\cal{N}}$ has real coefficients.  In particular,
\begin{align}
{\cal{N}}(z)&=8\sin(\gamma\tau)\sin^2(\gamma\tau/2)(1+2z\cos(\gamma\tau)+z^2);\nonumber\\
{\cal{D}}(z) &= -2[z^3(2\cos(\gamma\tau)+\cos(2\gamma\tau)) - 3 z^2(1+\cos(\gamma\tau)+\cos(3\gamma\tau)) 
+z(4\cos(\gamma\tau) + 5\cos(2\gamma\tau)) - 3].
\end{align}
Given this, we have
\begin{equation}
{\cal{F}} \equiv 1 - S_0 = \oint \frac{dz}{2\pi i z} \hat{\phi}^*(1/z)\hat{\phi}(z) = \oint \frac{dz}{2\pi i} \frac{\hat{\cal{N}}(1/z)\hat{\cal{N}}(z) }{D_0^2 \prod_i [(z-z_i)(z-1/z_i)]}.
\end{equation}
We can formally do the contour integral,
 picking up the residues at the three poles inside the unit circle,
 namely $1/z_i$, $i=1,\ldots,3$.  The result can be written as the ratio of polynomials in the three values $z_i$.  The key here is that both polynomials are {\em symmetric} under permutations of the three $z_i$'s.  Therefore, by the fundamental theorem of symmetric polynomials \cite{BenFin} (p. $90$),
each can be expressed uniquely in terms of the three elementary symmetric polynomials, $s_1=z_1+z_2+z_3$, $s_2=z_1z_2 + z_1z_3 + z_2z_3$ and $s_3=z_1z_2z_3$.  These three elementary polynomials are however simply related to the coefficients of the polynomial ${\cal{D}}$:
\begin{align}
s_1 &= 3 (1+\cos(\gamma\tau)+\cos(3\gamma\tau))/D_0; \nonumber\\
s_2 &= -(4\cos(\gamma\tau) + 5\cos(2\gamma\tau))/D_0; \nonumber\\
s_3 &= -3/D_0,
\end{align}
where
\begin{equation}
D_0 =  -2(2\cos(\gamma\tau)+\cos(2\gamma\tau))
\end{equation}
is the coefficient of the $z^3$ term in ${\cal{D}}$.  Performing these substitutions (via the command SymmetricReduction in Mathematica) in the numerator and denominator and simplifying, we find
${\cal{F}} =1$ so the survival is zero $S_0=0$.
This equation holds at all but the exceptional points,
  which do not have three poles in $\hat{\phi}(z)$ due to the collision of a pole (or pair of conjugate poles) with the zeros of ${\cal{N}}$ on the unit circle, leaving these cases to be examined individually.  Since ${\cal{D}}$ does not depend on the initial condition, the set of exceptional points is the same as for the $0\to 0$ transition.

This same general procedure works for the calculation of $\langle n \rangle$ as well, since again we only have the simple poles from ${\cal{D}}(1/z)$ to contend with, again giving three contributions.  The result of this exercise is as given in the main text.

\subsection{$L=6$, $1 \to 0$}
The same general procedure can be applied to the calculation of ${\cal{F}}$ for the $1$ to $0$ or $2$ to $0$  transitions,
 and gives ${\cal{F}}=1/2$ at all but the exceptional points, which again have to be handled separately.

\section{Calculating the integral, Eq. (\ref{eqGen01})}
\label{AppSolve}

The integration path in Eq.
(\ref{eqGen01})
 is shown in Fig. \ref{fig1path}
and since the integration along the outer zone $|z|\rightarrow \infty$ 
does not contribute,
we need to consider four integration segments, which are at a distance 
$\epsilon$ above and below the two branch cuts.   
We distinguish these four paths with indices $\sigma$ and $\beta$ that get
values $\pm 1$. $\sigma$ is an indicator for the branch cut,
$\sigma=+1$ represents the upper branch cut (see Fig. \ref{fig1path})
and $\sigma=-1$ the lower one. The index $\beta$ is for the direction
of integration,
$\beta=+1$ for outward integration in the radial direction while
$\beta=-1$ is for inward integration  (see Fig. \ref{fig1path}).
In the complex plane the parametrization of the four paths is 
\begin{equation}
z(y) = \left( 1 + y + i \epsilon \beta \right) e^{2 i \sigma \gamma \tau} ,
\ \ \  0<y<\infty \ \ \  \epsilon\rightarrow  0^{+}.
\label{eqGen03}
\end{equation}
Along these paths it is easy to show that
\begin{equation}
I_{\gamma \tau} \left[z(y)\right] ={1 \over 2 \sqrt{ \pi \gamma \tau}} 
\left\{
e^{ - i \sigma\pi/4} \mbox{Li}_{1/2} \left[\left( 1 + y + i  \epsilon \beta\right) e^{4 i \sigma \gamma \tau} \right]
+ e^{i \pi \sigma/4} \mbox{Li}_{1/2} \left(1 + y + i \epsilon \beta\right) \right\}.
\label{eqGen04}
\end{equation}
In the integration we consider the small $y$ limit
corresponding to  large $n$ 
using Eq. 
(\ref{eq48zeta}) along the four paths. The second term in
Eq. (\ref{eqGen04})
$\mbox{Li}_{1/2} \left(1 + y + i \epsilon \beta\right)\simeq \beta i \sqrt{\pi/y}$ is the large term and we get
\begin{equation}
\lim_{\epsilon \to 0} I_{\gamma \tau}[z(y)]\sim
{i \beta e^{  i \pi \sigma/4} \over 2 \sqrt{ \gamma \tau y} }
\label{eqGen05}
\end{equation}
The generating function is given by
\begin{equation}
\hat{\phi}\left[z(y)\right] \sim 1 - {1 \over I_{\gamma \tau}\left[z(y)\right]} \sim
1+ 2 i \beta e^{ -i \pi \sigma /4} \sqrt{ \gamma \tau y} 
\label{eqGen06}
\end{equation}
Integrating along the four lines, taking into consideration the clockwise
direction of the integration
we find
\begin{equation}
\phi_n =  \sum_{\beta=\pm,\sigma=\pm}
{\beta {\cal I}_{\sigma \beta} \over 2 \pi i},
\label{eqGen07}
\end{equation}
where
\begin{equation}
{\cal I}_{\sigma \beta} \sim  \int_0 ^\infty \underbrace{\exp\left\{
- \left( 1 + n \right) \ln \left[ \left( 1 + y + i \epsilon \beta\right) e^{i 2 \sigma \gamma \tau} \right] \right\} }_{z(y)^{-n-1}} 
\underbrace{\left(1 + 2 i \beta e^{ - i \sigma \pi/4} \sqrt{ \gamma \tau y} \right)}_{ \sim \hat{\phi}[z(y)]} \underbrace{e^{2 i \sigma \gamma \tau} {\rm d} y }_{{\rm d} z} .
\label{eqGen08}
\end{equation} 
In the limit $\epsilon \to 0$ the integration gives
\begin{equation}
{\cal I}_{\sigma \beta} \sim
e^{ - 2 i n \sigma \gamma \tau} \left( {1 \over n} + i \beta e^{ - i \sigma\pi/4} \sqrt{ 
\pi \gamma \tau \over n^{3}} \right)  
\label{eqGen09}
\end{equation} 
where we used $(1+n) \ln(1+y) \sim n y$ since $n$ is large
 and $\int_0 ^\infty \sqrt{y} \exp(-y n){\rm d} y= n^{-3/2} \sqrt{\pi}/2$. 
Using 
Eq. (\ref{eqGen07}) we find 
Eq. (\ref{eqGen10}).

\end{widetext}

\section{Survival probability for a particle on a line}
\label{AppD}

\begin{figure}
\centering
\includegraphics[width=0.49\textwidth]{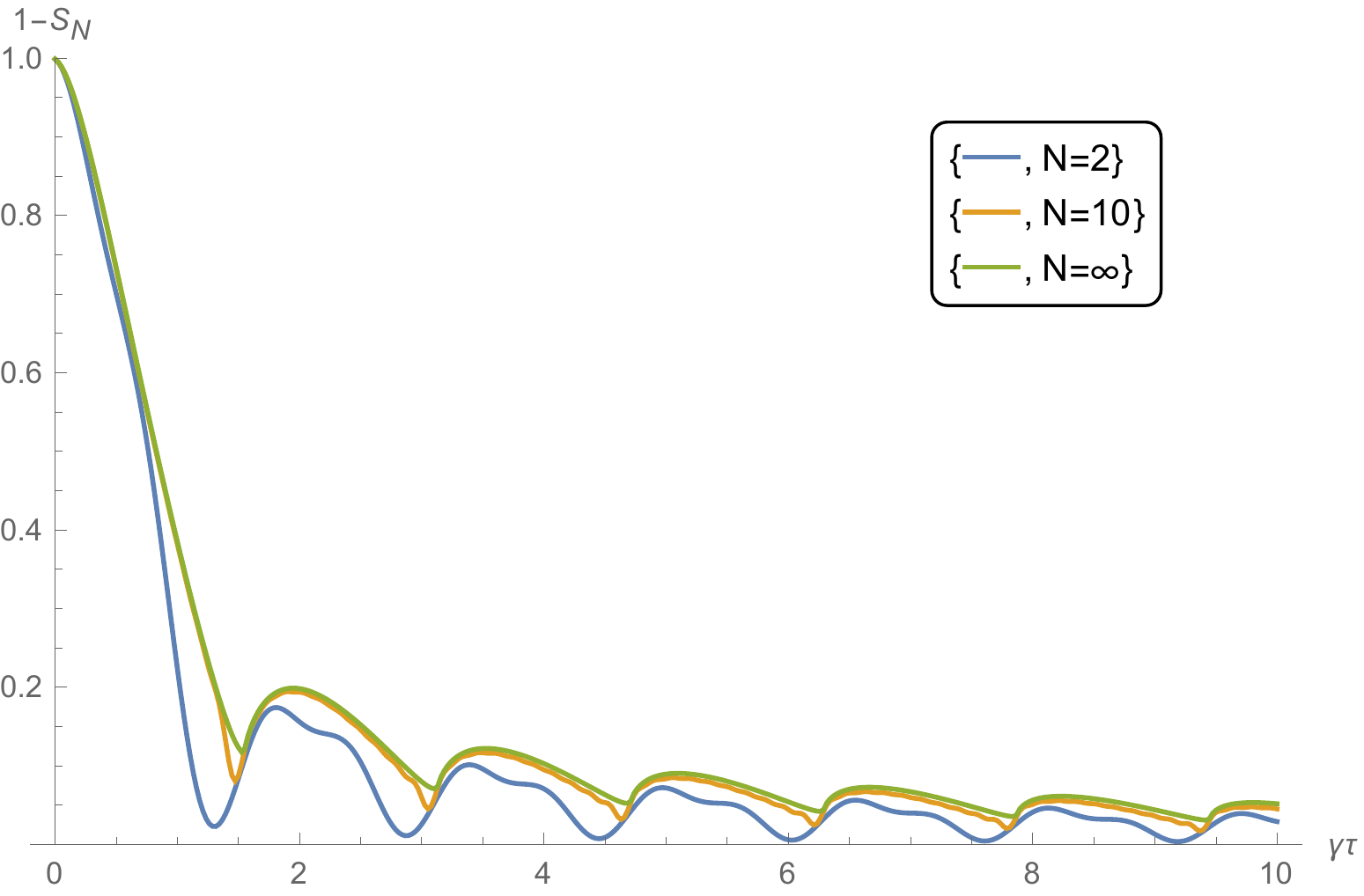}
\caption{
$1- S_N$ versus $\gamma \tau$ for a quantum walk on a line,  the
particle is launched from the origin. 
}
\label{figSNappD}
\end{figure}

 We presented the final detection probability $1 - S_\infty$ in Fig. 
\ref{figFGInf} for a particle starting on the origin of an infinite
line. Here we discuss briefly approximations for this probability.
A simple approximation is to consider the finite sum
\begin{equation}
1 - S_N = \sum_{n=1} ^N F_n .
\label{appDeq01}
\end{equation}
The values of $F_n$ are taken from Table 
\ref{Table01}.
As shown in Fig. 
\ref{figSNappD}
 already for $N=2$ the general features,
i.e. non monotonic decay of $1-S_\infty$ and periodic minima as
$\tau$ is varied  are clearly observed. 
This approximation works very well already for $N=10$. 
This shows that the small $n$ behavior of $F_n$ controls the
final survival probability. 

\begin{figure}
\centering
\includegraphics[width=0.49\textwidth]{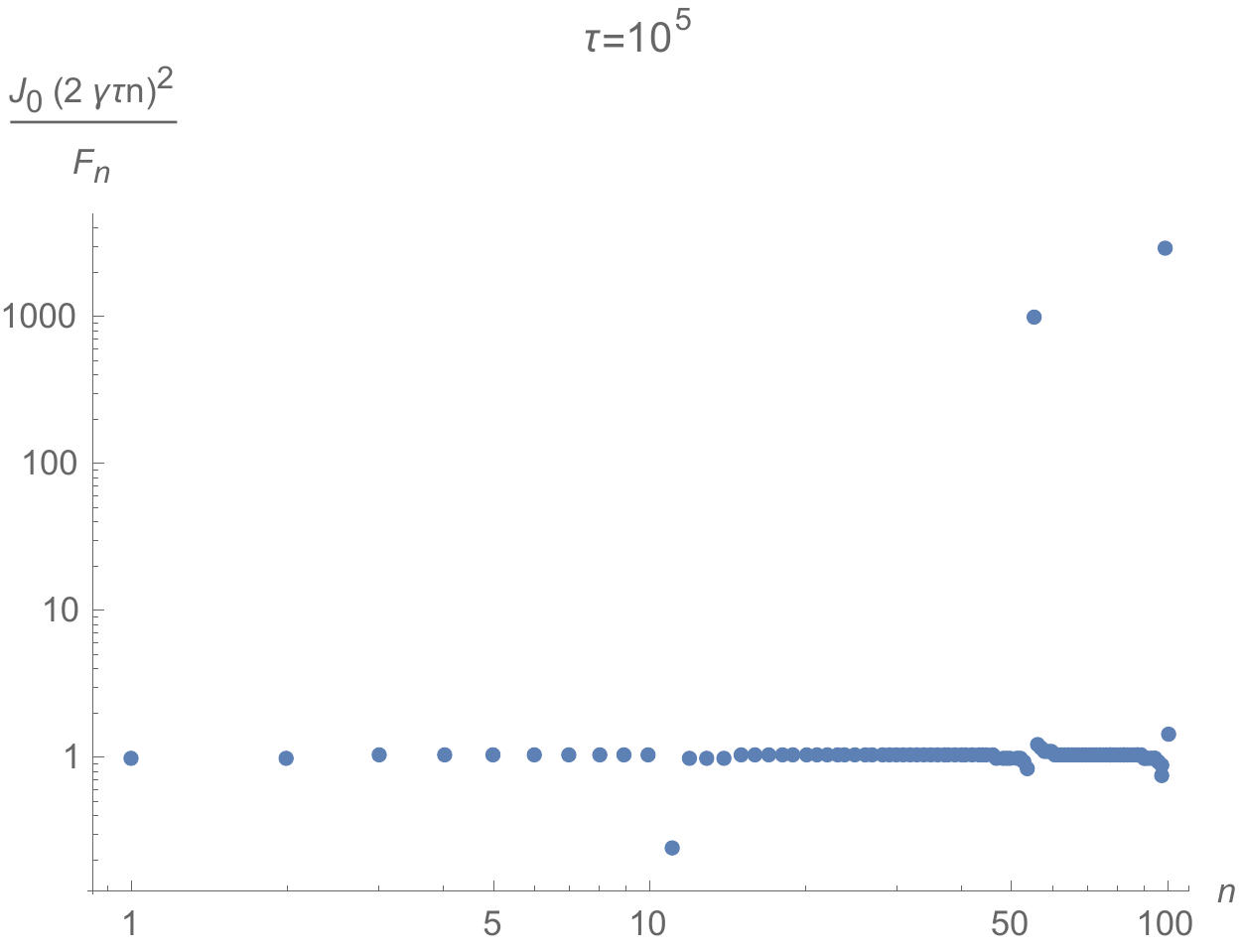}
\caption{
The ratio $J_0 \left( 2 n \gamma \tau\right)^2 / F_n$ for $\gamma \tau =10^5$. 
}
\label{figAppDcom0}
\end{figure}

\begin{figure}
\centering
\includegraphics[width=0.49\textwidth]{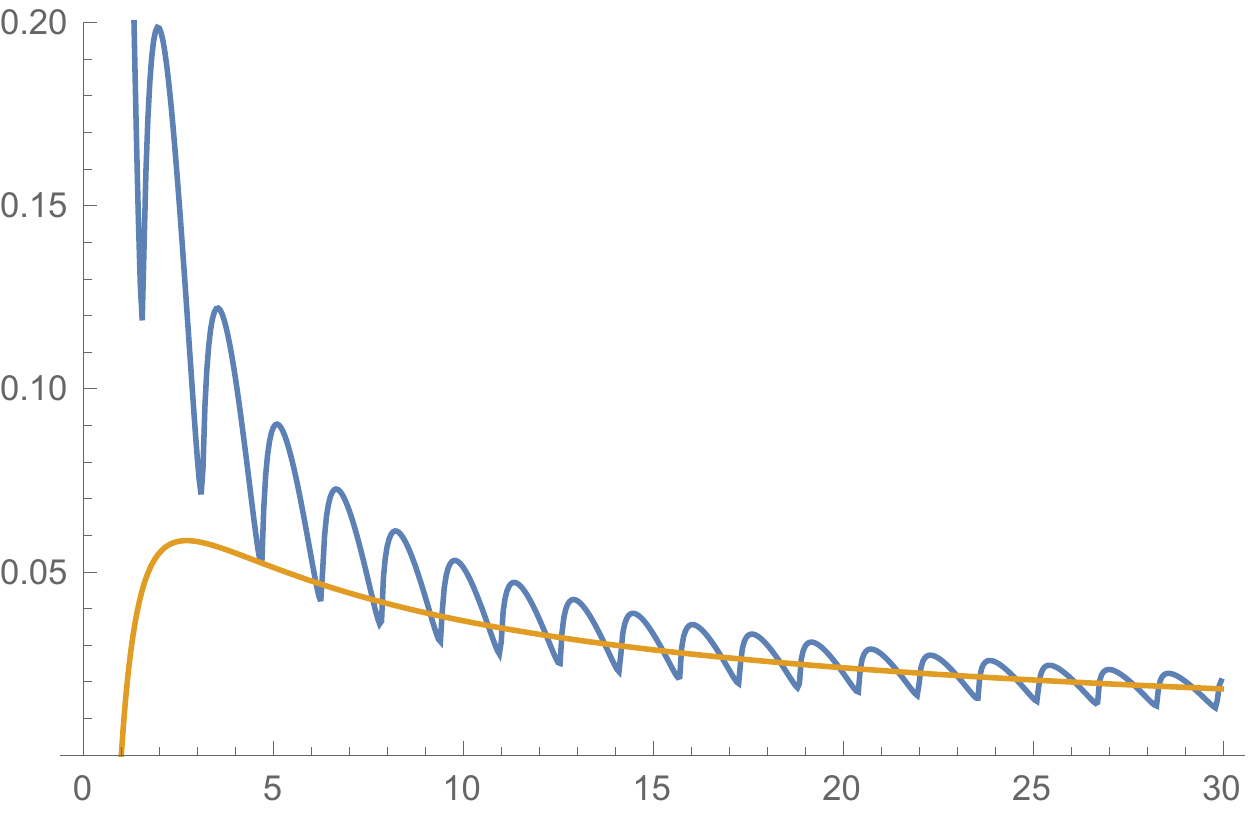}
\caption{
$1- S_\infty$ versus $\gamma \tau$. The approximation Eq. 
(\ref{eqSUMINT}) is compared with the exact result, and it works reasonably 
well for large $\gamma \tau$, as expected. However, it does not predict the
spiky cusps neither the non monotonic behavior of $1-S_\infty$. 
}
\label{figAppDcom}
\end{figure}

\begin{figure}
\centering
\includegraphics[width=0.49\textwidth]{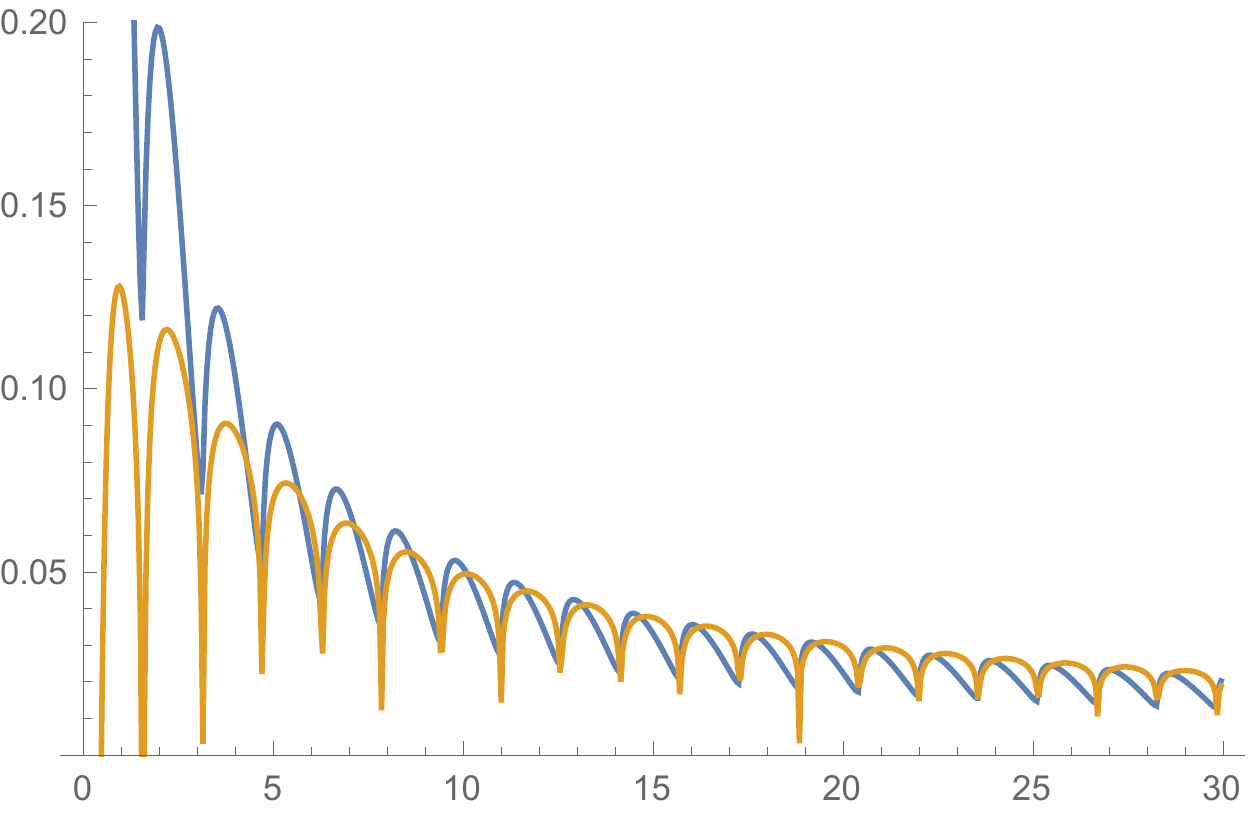}
\caption{
$1- S_\infty$ versus $\gamma \tau$. The approximation Eq. 
(\ref{eqKestheta}) works well. 
}
\label{figAppDcom1}
\end{figure}

 For large $\gamma \tau$ we have $\phi_n \simeq J_0 \left( 2 n \gamma \tau\right)$ [see discussion above  Eq. 
(\ref{eqPPHH})]. This approximation is compared with the exact result in Fig.
\ref{figAppDcom0}. In this limit of large $\gamma \tau$ 
Eq.  
(\ref{eqPPHH})
holds. The approximation for $1- S_\infty$ Eq.  
(\ref{eqSUMINT}) which also works in the large $\gamma \tau$ limit 
is tested in Fig. 
\ref{figAppDcom}. The approximation is just qualitative. 
We obtained a far  better approximation 
\begin{equation}
1 - S_\infty \simeq {1 \over 4 \pi \gamma \tau} \left\{ 
\ln \left[ 16 \pi^2 (\gamma\tau)^2 \theta^{*} \left( \pi - \theta^{*} \right) \right] - 2 \right\},
\label{eqKestheta}
\end{equation}
where $\theta^{*} =\mbox{modulo} \left( 2 \gamma \tau , \pi \right)$. 
This approximation is demonstrated in Fig. 
\ref{figAppDcom1}.

\begin{thebibliography}{99}
\bibitem{remark1} 
 Note that in textbooks
the $Z$ transform is defined as $f(z)=\sum_{n=0} ^\infty f_n z^n$ while
in our case the first term $n=0$ is zero, i.e. $\phi_0=0$ (since
the first measurement is at time $\tau$). Consistently
Eqs. (\ref{eq21},\ref{eq22}) give $\hat{\phi}(z)|_{z=0} = 0$.
\end{thebibliography}
\end{document}